\documentclass[structabstract]{aa}  
\usepackage{graphicx}
\usepackage{natbib}
\bibpunct{(}{)}{;}{a}{}{,} % to follow the A&A style
\usepackage{ulem}
\newcommand{\sqdeg}{\,deg$^2$}
\usepackage{txfonts}
\begin{document}
\title{The ultracool-field dwarf luminosity-function and space density from the 
Canada-France Brown Dwarf Survey
\thanks{Based on observations obtained with MegaPrime/MegaCam, a joint
  project of CFHT and CEA/DAPNIA, at the Canada-France-Hawaii
  Telescope (CFHT), which is operated by the National Research Council
  (NRC) of Canada, the Institut National des Sciences de l'Univers of
  the Centre National de la Recherche Scientifique (CNRS) of France,
  and the University of Hawaii. This work is based in part on data
  products produced at TERAPIX and the Canadian Astronomy Data Centre
  as part of the Canada-France-Hawaii Telescope Legacy Survey, a
  collaborative project of NRC and CNRS. 
  Based on observations made with the
ESO New Technology Telescope at the La Silla Observatory.
% under programme ID 76.C-0540(A), 77.C-0594, 77.A-0707,  78.A-0651,78.C-0629 and 79.A-0663. 
Based on observations  obtained  at the Gemini Observatory,
  which is operated by the Association of Universities for Research in
  Astronomy, Inc., under a cooperative agreement with the NSF on
  behalf of the Gemini partnership: the National Science 
Foundation (United States), the Science and Technology Facilities
Council (United Kingdom), the National Research Council (Canada),
CONICYT (Chile), the Australian
Research Council (Australia), CNPq (Brazil) and CONICET (Argentina).
Based on
observations with the Kitt Peak National Observatory, National Optical
Astronomy Observatory, which is operated by the Association of
Universities for Research in Astronomy, Inc. (AURA) under cooperative
agreement with the National Science Foundation. 
Based on observations
made with the Nordic Optical Telescope, operated on the island of La
Palma jointly by Denmark, Finland, Iceland, Norway, and Sweden, in the
Spanish Observatorio del Roque de los Muchachos of the Instituto de
Astrofisica de Canarias. 
Based on observations made at The McDonald Observatory of the University of Texas at
Austin.}
}

%   \subtitle{I. Overviewing the $\kappa$-mechanism}
\titlerunning{Ultracool-field dwarf luminosity-function and space density from the 
CFBDS}

\author{C. Reyl\'e \inst{1}
  \and P. Delorme \inst{2}
  \and C.J. Willott \inst{3}
  \and L. Albert \inst{4}
  \and X. Delfosse \inst{5}
  \and T. Forveille \inst{5}
  \and E. Artigau \inst{6} 
  \and L. Malo \inst{6}
  \and G.J. Hill\inst{7}
  \and R. Doyon \inst{6}
 }

% \offprints{C\'eline Reyl\'e, \email{celine@obs-besancon.fr}}

\institute{Observatoire de Besan\c{c}on, Universit\'e de Franche-Comt\'e, Institut Utinam, UMR CNRS 6213, 
   BP 1615, 25010 Besan\c{c}on Cedex, France\\
  \email{celine@obs-besancon.fr}
  \and School of Physics and Astronomy, University of St Andrews, North Haugh, St Andrews KY16 9SS, United Kingdom  
  \and Herzberg Institute of Astrophysics, National Research Council, 5071 West Saanich Rd, Victoria, BC V9E 2E7, Canada.
  \and Canada-France-Hawaii Telescope Corporation, 65-1238 Mamalahoa 
   Highway, Kamuela, HI96743, USA
  \and  Laboratoire d'Astrophysique de Grenoble,Universit\'e
  J.Fourier, CNRS, UMR5571, Grenoble, France
%  \and Gemini Observatory Southern Operations Center c/o AURA, Casilla 603
%   La Serena, Chile
  \and D\'epartement de physique and Observatoire du Mont M\'egantic,
  Universit\'e de Montr\'eal, C.P. 6128, Succursale Centre-Ville,
  Montr\'eal, QC H3C 3J7, Canada
  \and McDonald Observatory, University of Texas at Austin, 1
  University Station C1402, Austin, TX 78712-0259, USA 
}

   \date{}

% \abstract{}{}{}{}{} 
% 5 {} token are mandatory
 
  \abstract
   {Thanks to recent and ongoing large scale surveys, hundreds of brown dwarfs have been discovered in the last decade. The Canada-France Brown Dwarf Survey is a wide-field survey for cool brown dwarfs conducted with the MegaCam camera on the Canada-France-Hawaii
  Telescope telescope.}
   {Our objectives are to find ultracool brown dwarfs and to
constrain the field brown-dwarf luminosity function 
and the mass function from a large and homogeneous sample of L and
T dwarfs.}
   {We identify candidates in CFHT/MegaCam $i'$ and $z'$ images and 
follow them up with pointed near infrared (NIR) imaging on several telescopes. Halfway through our survey we found $\sim$50 T dwarfs and $\sim$170 L or ultra cool M dwarfs drawn from a larger sample of 
1400 candidates with typical ultracool dwarfs $i'-z'$ colours, found in 780 square degrees.}
   {We have currently completed the NIR follow-up on a large part of the survey for all candidates from mid-L dwarfs down to the latest T dwarfs known with  utracool dwarfs' colours. This allows us to draw on a complete and well defined sample of 102 ultracool dwarfs to investigate the luminosity function and space density of field dwarfs. }
   {We found the density of late L5 to T0 dwarfs to be $2.0^{+0.8}_{-0.7} \times 10^{-3}$ objects pc$^{-3}$, the density of T0.5 to T5.5 dwarfs to be $1.4^{+0.3}_{-0.2} \times 10^{-3}$ objects pc$^{-3}$, and the density of T6 to T8 dwarfs to be $5.3^{+3.1}_{-2.2} \times 10^{-3}$ objects pc$^{-3}$.  We found that these results agree better with a flat substellar mass function.
   Three latest dwarfs at the boundary between T and Y dwarfs give the high density $8.3^{+9.0}_{-5.1} \times 10^{-3}$ objects pc$^{-3}$. Although the uncertainties are very large this suggests that many brown dwarfs should be found in this late spectral type range, as expected from the cooling of brown dwarfs, whatever their mass, down to very low temperature. }


   \maketitle
%
%________________________________________________________________

\section{Introduction}

Brown dwarfs are very low-luminosity objects. Even at their flux maximum in the near and mid-infrared, brown dwarfs are more than ten magnitudes fainter than solar-type stars. 
That explains the very low number of detected substellar objects compared to that of known stars, although they probably represent a sizeable fraction of the stellar population in our Galaxy.

%Their detection is almost as tricky as that of an exoplanet and the first discovery of these two kind of objects have been made the same year \citep{Nakajima.1995,Mayor.1995} and today the number of known brown dwarfs is only twice that of exoplanets. 
The first discoveries of brown dwarfs were made by \cite{Nakajima.1995} around the nearby star Gliese 229 and by 
\cite{Stauffer.1994} and \cite{Rebolo.1995} in the Pleiades.
\cite{Delfosse.1997,Ruiz.1997} and \cite{Kirkpatrick.1997} found the first field
brown dwarfs.
Since then, several hundreds of field brown dwarfs have been identified, most of them thanks to large-scale optical and near-infrared imaging surveys because they can be identified by their red optical minus near-infrared colours. Most of them have been found in the Two-Micron All-Sky Survey \citep[2MASS,][]{Skrutskie.2006} and in the Sloan Digital Sky Survey \citep[SDSS,][]{York.2000}. See \cite{Kirkpatrick.2005} for a review on brown dwarf discoveries.

The new generation of large-area surveys uses deeper images.
As a consequence, the number of known brown dwarfs increases and new types of rarer and fainter brown dwarfs are detected \citep{Delorme.2008a,Burningham.2008}. Such surveys are the UKIRT Infrared Deep Sky Survey \citep[UKIDSS, ][]{Lawrence.2007} and
 the one we undertook, the Canada-France-Brown-Dwarf Survey \citep[CFBDS, ][]{Delorme.2008b}. The Canada-France-Brown-Dwarf Survey is based on deep multi-colour MegaCam optical imaging obtained at the Canada-France-Hawaii Telescope (CFHT). We expect that complete characterisation of all our candidates will yield about 100 T dwarfs and over 400 L or very late-M dwarfs, which will approximately double the number of known brown dwarfs with a single, well-characterised survey. 

With that large number of identified brown dwarfs, it becomes possible to define uniform and well-characterised samples of substellar objects to investigate their mass and luminosity functions. 
The knowledge of these functions is essential in several studies. In terms of Galactic studies, it gives clues on the baryonic content of the Galaxy and contributes to determine the evolution of the Galaxy mass. In terms of stellar physics studies, it gives constraints on stellar and substellar formation theories.

It has been shown that a single log-normal function could fit the mass function from field stars to brown dwarfs \citep[see][]{Chabrier.2003}).
\citep[see also][]{Luhman.2007b}). Our Galaxy probably counts several $10^{10}$ of brown dwarfs, hundreds of them in the Solar neighbourhood!

To validate these assumptions, which are mainly based on the study of brown dwarfs in stars clusters, it is necessary to first refine the field brown dwarf luminosity-function. An initial estimate of the local space density of late T dwarfs have been made by \cite{Burgasser.2002PhD}\footnote{http://web.mit.edu/~ajb/www/thesis}, from a sample of 14 T-dwarfs.  \cite{Allen.2005} used these results combined with data compiled from a volume-limited sample of late M and L dwarfs  \citep{Cruz.2003} to compute a luminosity function.
Recently, a detailed investigation on a volume-limited sample of field late-M and L dwarfs has been performed by \cite{Cruz.2007}. A similar empirical investigation of field T-dwarfs have been presented by \cite{Metchev.2008}. Still the number of objects in their sample is relatively small (46 L-dwarfs and 15 T-dwarfs, respectively) and the field brown dwarf luminosity-function remains poorly constrained. Thus further efforts are needed to measure the space density of brown dwarfs. 

At mid-course of the CFBDS survey, we are able to define an homogeneous sample of 102 brown dwarfs redder than $i'-z'=2$, from the mid-L dwarfs to the far end of the brown dwarfs observed sequence at the T/Y transition, and to derive a luminosity function. These objects are drawn from a 444 square degree 
area (57\% of the total area) where all candidates with $i'-z'>2.0$  have been followed-up 
with near-infrared photometry and, for the reddest of them, with spectroscopy. They represent a sub-sample in colour ($i'-z'<2.0$) and in 
magnitude ($z'<22.5$) of the hundreds of brown dwarfs found with $i'-z'>1.7$ on the entire survey.
Sect.~\ref{obs} briefly describes the CFBDS and the related observations. The construction of a complete and clean sample of L5 and cooler dwarfs is explained in Sect.~\ref{sample}. In Sect.~\ref{disc}, we compare our results with prior studies and link them to the mass function of brown dwarfs. Sect.~\ref{lf} presents the luminosity function derived from this sample. Conclusions are given in Sect.~\ref{ccl}.
%__________________________________________________________________

\section{Observations}
\label{obs}

The brown dwarf sample is drawn from the CFBDS. The survey is fully described in \cite{Delorme.2008b}.
The CFBDS is a survey in the $i'$ and $z'$ filters
conducted with MegaCam \citep{Boulade.2003proc} at the CFHT. 
It is based on two existing surveys, the CFHT Legacy Survey \citep[CFHTLS\footnote{http://www.cfht.hawaii.edu/Science/CFHLS/},][]{cuillandre.2006} and the Red-sequence Cluster Survey \citep[RCS-2,][]{yee.2007}, complemented with significant Principal Investigator data at CFHT. 
The survey is also extremely effective at finding high-redshift quasars. The parallel programme is called the Canada-France High-z Quasar Survey, and results are presented in \cite{Willott.2007,Willott.2009}.

The $i'$ and $z'$ imaging part of the 780 \sqdeg\  CFBDS is nearing completion to typical limit of $z'=22.5$, probing the brown dwarf content of the Galaxy as far as 215 pc for the mid-L dwarfs, 180 pc for the early-type T dwarfs and 50 pc for the late-type T dwarfs.

The reddest sources are then followed-up with pointed $J$-band imaging to distinguish brown dwarfs from other astronomical sources, 
and spectra are obtained for the latest type dwarfs.

Throughout this paper, we use Vega magnitudes for the $J$-band and AB magnitudes for the $i'$ and $z'$ optical bands.

\subsection{Optical imaging}

We only briefly describe the processing and photometry of the
imaging observations in this paper. A full description
can be found in \cite{Delorme.2008b}. 

Pre-processing of
the MegaCam images is carried out at CFHT using the ELIXIR
pipeline. This removes the instrumental effects from the images.
We then run our own algorithms to improve the astrometry and
check the photometry. Finally, we stack the images (if there is more than one exposure at a given position)
and register the images in the four different filters. Photometry
is carried out with an adaptation of SExtractor \citep{Bertin.1996}, 
which uses dual-image multiple point-spread function fitting to
optimise the signal-to-noise ratio (S/N) of point sources.

Candidate brown dwarfs and quasars are initially identified on the optical images
as objects which have very high $i'-z'$ colours. A 10 $\sigma$ detection in $z'$ is required but no constraint is set on $i'$: in several cases, the targets are $i'$-dropouts and only $i'-z'$ lower limits can be determined. As shown in \cite{Willott.2005}, photometric noise causes many
M dwarfs with intrinsic colours of $0.5< i' - z' <1.5$ to be scattered
into the region of the diagram at $i'-z' >1.5$ where we would
expect to find only L or T dwarfs and quasars. The huge number
of these M stars would require a lot of telescope time for complete
follow-up. Therefore, we limit our survey to objects
observed to be redder than $ i'-z' >1.7$. Changing this criterion
reduces the number of M dwarf contaminants by $\sim$ 95\%.

\subsection{Near-infrared imaging}

As shown by \cite{fan.2001}, brown dwarfs and quasars (and artefacts) can be
separated with NIR $J$-band imaging. \cite{Willott.2005} described in detail 
the method for identifying high-redshift quasars and brown dwarfs using MegaCam optical 
plus near-IR imaging. 
The very red $i'-z'$ of high-redshift quasars is caused by deep 
Lyman-$\alpha$ absorption on a relatively flat intrinsic spectrum, and they
appear significantly bluer than brown dwarfs in $z'-J$. On the contrary the spectral distribution of brown dwarfs keeps 
rising into the $J$ band. We therefore carried out NIR imaging at several observatories: 
La Silla (New Technology Telescope, 3.6m), McDonald (2.7m), Kitt Peak (2.1m), La Palma (Nordic Optical Telescope, 2.5m). 
We obtained the $J$ magnitude for all candidates, dwarfs or
quasars. For dwarfs, the signal-to-noise ratio is 10 to 50 $\sigma$.

Besides pinpointing the few high-redshift quasars that contain important clues on the reionization of the Universe 
\citep{Willott.2005,Willott.2007,Willott.2009}, the $J$-band photometry very effectively rejects any remaining observational 
artefacts, as well as the more numerous M stars
scattered into the brown dwarf/quasar box by large noise excursions.

\subsection{Spectroscopy}

Follow-up spectroscopy of the T dwarf candidates with the reddest $z'-J$ colour
was then carried out at Gemini. Before the
GNIRS incident at Gemini-South on April 2007 (see Gemini Observatory call for proposal archive, semester 2007B\footnote{http://www.gemini.edu/sciops/observing-with-gemini/previous-semesters/cfp-archive?q=node/11034}), it was used in its
cross-dispersed mode to obtain a 0.9-2.4 micron coverage. Then NIRI at
Gemini-North was used in a two-step approach: 1) $H$-band spectra were obtained
to determine spectral types and 2) T6 and cooler objects were targeted for
additional coverage in the $J$ band and occasionally in the $K$ band.

To date we obtained spectroscopic
follow-up for 40 T-dwarfs. Four more T-dwarf candidates have been followed-up by \cite{Knapp.2004,Chiu.2006,Warren.2007}. Spectral types range between T0 and the T/Y transition with 6 dwarfs being of type T7 or later.
Four of these have already been presented in
\citep{Delorme.2008a,Delorme.2008b}. Spectral indices of the others are given in Table~\ref{tab_spectype} and are fully described in a forthcoming paper \citep[see also][]{Albert.2009}. For all of them we use an absolute magnitude
versus spectral type relation instead of the absolute magnitude versus colour
used for the other objects. (see Sect.~\ref{absmag}).

\begin{table*}

\caption{Spectral indices for the the unified scheme of
\citet{Burgasser.2006} and spectrophotometric distances. Each index translates
to a spectral type (between brackets) through linear interpolation using
\citeauthor{Burgasser.2006}'s Table 5. The adopted spectral type is a straight
average of the spectral indices that could be measured (upper limits are
rejected). 
\label{tab_spectype}}

%\scriptsize
\center

\begin{tabular}{rcccccc}

\hline\hline

{Designation} & {H$_2$O$-J$} & {CH$_4-J$} & {H$_2$O$-H$} & {CH$_4-H$} & {CH$_4-K$} &{Adopted SpT}  \\

\hline

%ULAS~J003402$-$005205    &0.012[$>$T8] &0.014[$>$T8] & 0.133[$>$T8] & 0.096[T8]  &0.091[$>$T7]    &T9.0     \\
CFBDS~J005910$-$011401 & 0.037[T8.66] &  0.175[T8.58] & 0.125[$>$T8] & 0.086[T8.81] & 0.111[$>$T6] & T9.0 \\
CFBDS~J025401$-$182529 & 0.160[T6.40] &  0.350[T6.12] & 0.267[T6.88] & 0.244[T7.06] & 0.163[T6.35] & T6.5 \\
CFBDS~J025558$-$173020 & ------------ &  ------------ & 0.580[T1.29] & 0.980[T0.66] & ------------ & T1.0 \\ 
CFBDS~J025718$-$124853 & ------------ &  ------------ & 0.412[T4.30] & 0.774[T3.13] & ------------ & T3.5 \\ 
CFBDS~J025805$-$145534 & 0.555[$<$T2] &  0.642[T2.32] & 0.568[T1.45] & 1.013[$<$T0] & 0.723[T1.22] & T1.5\\
CFBDS~J025840$-$182648 & ------------ &  ------------ & 0.559[T1.59] & 0.702[T3.49] & ------------ & T3.0 \\ 
CFBDS~J030130$-$104504 & ------------ &  ------------ & 0.325[T5.90] & 0.378[T5.86] & ------------ & T6.0 \\
CFBDS~J030225$-$144125 & ------------ &  ------------ & 0.306[T6.23] & 0.434[T5.38] & ------------ & T5.5 \\
CFBDS~J030226$-$143719 & 0.381[T3.99] &  0.595[T2.83] & 0.408[T4.37] & 0.563[T4.31] & ------------ & T4.0 \\
CFBDS~J090139$+$174051 & ------------ &  ------------ & 0.444[T3.55] & 0.630[T3.85] & ------------ & T4.0 \\ 
CFBDS~J090449$+$165347 & ------------ &  ------------ & 0.408[T4.37] & 0.596[T4.03] & ------------ & T4.0 \\
CFBDS~J092250$+$152741 & ------------ &  ------------ & 0.279[T6.68] & 0.238[T7.12] & ------------ & T7.0 \\ 
CFBDS~J102841$+$565401 & 0.043[T8.53] &  0.257[T7.34] & 0.179[T8.35] & 0.147[T8.04] & ------------ & T8.0 \\
CFBDS~J104209$+$580856 & 0.160[T6.40] &  0.349[T6.16] & 0.248[T7.21] & 0.269[T6.83] & 0.167[T6.25] & T6.5 \\
CFBDS~J145044$+$092108 & ------------ &  ------------ & 0.483[T2.68] & 0.690[T3.55] & ------------ & T3.5\\
CFBDS~J145847$+$061402 & ------------ &  ------------ & 0.552[T1.69] & 0.871[T2.41] & ------------ & T2.0 \\
CFBDS~J145935$+$085751 & ------------ &  ------------ & 0.405[T4.42] & 0.528[T4.60] & ------------ & T4.5\\
CFBDS~J150000$-$182407 & 0.335[T4.89] &  0.441[T5.10] & 0.387[T4.72] & 0.485[T4.96] & 0.429[T3.61] & T4.5 \\
%CFBDS~J150210.19$+$035055.51                                                                                                       $309\pm 56$\\
% SDSS~J150411$+$102717                                                                                                              $18\pm  3$\\
%SDSS~J151114$+$060742 & 0.619[$<$T2] & 0.649[T2.23]  & 0.653[T0.12] & 0.802[T2.99] & 0.865[$<$T0] & T2.0\\ 
CFBDS~J151803$+$071645 & ------------ &  ------------ & 0.535[T1.93] & 0.788[T3.06] & ------------ & T2.5\\
CFBDS~J152514$+$111833 & ------------ &  ------------ & 0.500[T2.43] & 0.833[T2.73] & ------------ & T2.5\\
CFBDS~J152655$+$034536 & 0.464[T2.86] &  0.532[T3.80] & 0.409[T4.35] & 0.574[T4.22] & 0.305[T4.41] & T4.0\\
CFBDS~J203737$-$192202 & ------------ &  ------------ & 0.642[T0.29] & 0.998[T0.08] & ------------ & T0.0\\ 
CFBDS~J203841$-$185012 & ------------ &  ------------ & 0.499[T2.45] & 0.781[T3.10] & ------------ & T3.0\\
CFBDS~J204803$-$183212 & ------------ &  ------------ & 0.381[T4.82] & 0.537[T4.53] & ------------ & T4.5 \\
CFBDS~J212243$+$042942 & ------------ &  ------------ & 0.571[T1.42] & 0.864[T2.46] & ------------ & T2.0\\
% SDSS~J212413$+$010002                                                                                                              $20\pm  4$\\
CFBDS~J212702$+$002344 & ------------ &  ------------ & 0.450[T3.32] & 0.732[T3.34] & ------------ & T3.5 \\ 
CFBDS~J214139$-$033739 & ------------ &  ------------ & 0.529[T2.02] & 0.994[T0.21] & ------------ & T1.0 \\
CFBDS~J223856$+$034947 & ------------ &  ------------ & 0.561[T1.55] & 0.848[T2.60] & ------------ & T2.0\\
\hline\hline

\end{tabular}

%\tablenotetext{a}{This is a tight binary with rough spectral types of T4 and T6, placing them at $\sim40$~pc (Liu, in preparation).}

%}

\end{table*}

\section{Defining a homogeneous, complete, and clean sample}

\subsection{Photometric classification}
\label{classification}

One square-degree MegaCam image contains several hundred thousand objects, of 
which at most a few are brown dwarfs. We thus need to strike a careful
balance between sample completeness and contamination. To tune this
compromise we need a precise knowledge of the colours of brown dwarfs 
for the exact instruments and filters used in our survey.

The spectral energy distribution obtained from several photometric bands is characteristic of
the spectrum of one object. In order to get information on the spectral type of the
CFBDS candidates based on its photometry, we determined the colours in the photometric system
used by CFBDS of brown dwarfs with known spectral type.

We used publicly available spectra from the L and T dwarf data archive\footnote{http://www.jach.hawaii.edu/$\sim$skl/LTdata.html},
\citep{Martin.1999,Kirkpatrick.2000,Geballe.2001,Leggett.2002,Burgasser.2003,Knapp.2004,Golimowski.2004,Chiu.2006}
of 45 brown dwarfs with spectral types L1 to T8. Spectral types for L and T dwarfs are given according to the infrared classification scheme described in \cite{Geballe.2002} and \citet{Burgasser.2006}, respectively.

Thus synthetic colours in the MegaCam filters are computed in the AB system \citep{Fukugita.1996} 
using detector quantum efficiency and transmission curves for the atmosphere, telescope, camera 
optics, and filters. 
We similarly synthesised $J$-band photometry for each of the 
instruments and $J$ filters used in the $J$-band follow up. These 
instruments have significantly different response curves, which must be taken into account to obtain homogeneous selection criteria. 

\begin{figure}[!h]
\begin{center}
\includegraphics[scale=0.45,angle=270,clip=]{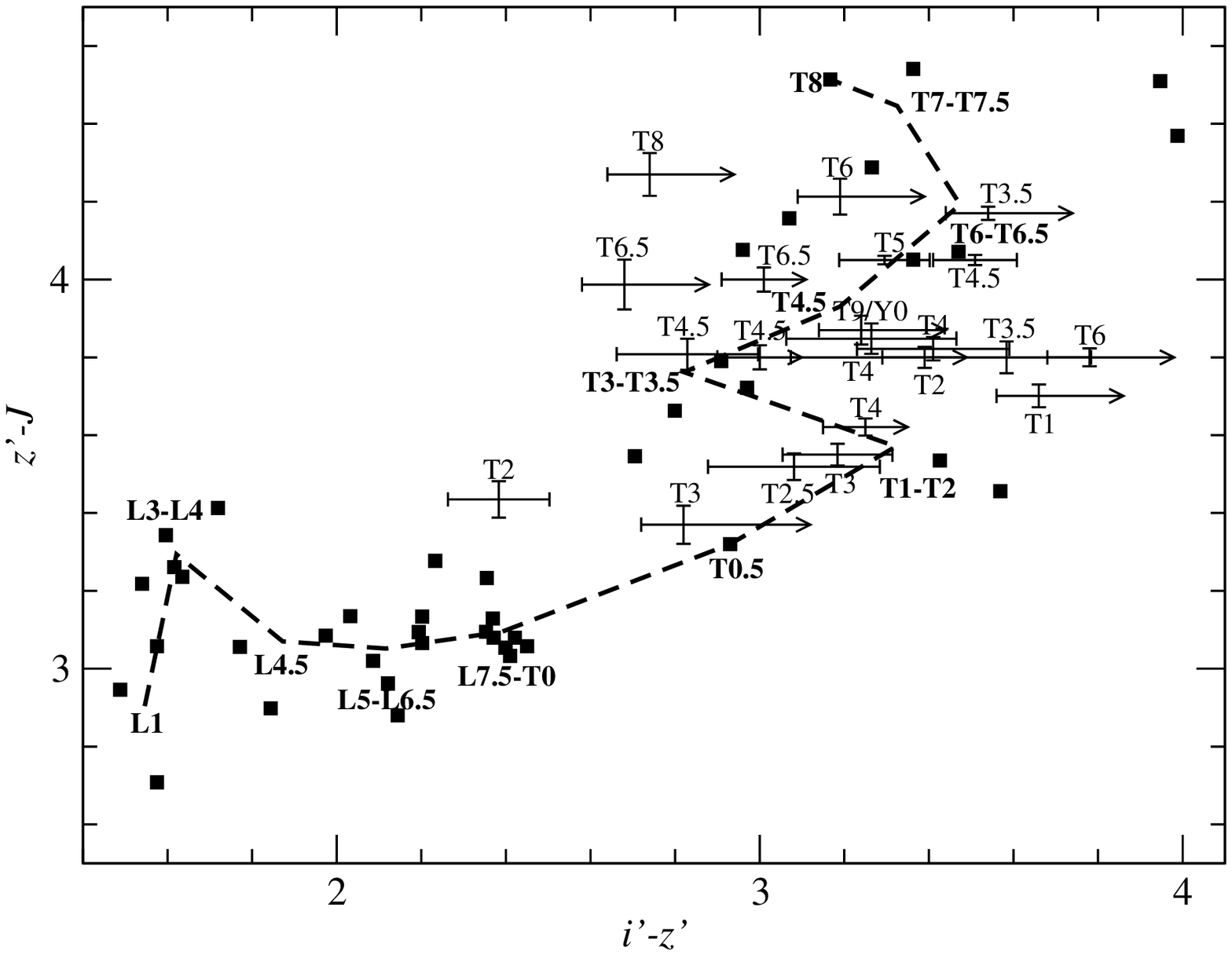}
\caption{
   $z'-J$ synthetic colour versus $i'-z'$ synthetic colour computed from available spectra in the literature. 
$i'$ and $z'$ are computed in the CFHT/MegaCam filters, $J$ is computed for the NTT/SOFI $J_s$ filter. The filled squares show the colours of each brown dwarf. The dashed line shows the mean colour-colour relation. Spectral types are given along this track, indicating the averaged colour of brown dwarfs with a given spectral type. Crosses with error bars show the T dwarfs for which we obtained spectroscopic observations. An arrow indicates no detection in the $i'$-band, meaning that the $i'-z'$ colour is a lower limit.
\label{iz-zJ_synth}
}
\end{center}
\end{figure}

Synthetic colours are given in Table~\ref{tabsynth}. Figure~\ref{iz-zJ_synth} displays the resulting colour-colour diagram. The synthetic colours of each brown dwarf are represented by filled squares. The dashed line shows 
the resulting colour-colour relation. Spectral types are given along this track, indicating the averaged colour of brown dwarfs at a given spectral type. 

\begin{table}[h]
\caption{Synthetic colours computed on publicly available spectra.}             
\label{tabsynth}      
\begin{tabular}{l l  l l l l l}        
\hline\hline                 
  Name		&spT		&$i'$		&$z'$	&$J$		&$i'$-$z'$	&$z'$-$J$\\
\hline                        
2MASS0345+25	&L1		&18.47	&16.89	&13.84	&1.58	&3.05\\
2MASS1439+19	&L1			&16.93	&15.36	&12.65	&1.57	&2.71\\
2MASS0028+15	&L3			&21.51	&19.87	&16.63	&1.64	&3.24\\
2MASS2224-01	&L3.5		&19.01	&17.29	&13.88	&1.72	&3.41\\			
DENIS1058-15	&L3			&18.87	&17.33	&14.12	&1.54	&3.22	\\	
GD165B			&L3			&20.51	&18.89	&15.63	&1.62	&3.26\\
2MASS0036+18	&L4				&17.25	&15.66	&12.31	&1.59	&3.35\\		
LHS102B			&L4.5		&17.82	&16.04	&12.99	&1.78	&3.05\\			
SDSS0835+19		&L4.5		&21.12	&19.14	&16.06	&1.98	&3.08\\		
2MASS1507-16	&L5.5		&17.44	&15.59	&12.69	&1.85	&2.90\\				
DENIS0205-11	&L5.5		&19.48	&17.36	&14.40	&2.12	&2.96	\\		
SDSS0107+00		&L5.5	&20.98	&18.78	&15.71	&2.20	&3.07\\			
SDSS0206+22		&L5.5		&20.88	&19.45	&16.42	&1.43	&3.03\\		
SDSS1326-00		&L5.5		&21.18	&19.03	&16.15	&2.15	&2.88\\			
SDSS0539-00		&L5			&18.95	&16.87	&13.85	&2.08	&3.02\\		
SDSS1446+00		&L5			&20.72	&18.69	&15.55	&2.03	&3.14	\\	
2MASS0825+21	&L6				&20.39	&18.16	&14.88	&2.23	&3.28	\\	
SDSS0236+00		&L6.5		&21.33	&19.13	&15.99	&2.20	&3.14	\\		
DENIS1228-15	&L6			&19.44	&17.57	&14.28	&1.87	&3.29		\\
2MASS1632+19	&L7.5		&21.25	&18.88	&15.75	&2.37	&3.13	\\		
2MASS2244+20	&L7.5			&21.65	&19.65	&16.32	&2.00	&3.33	\\		
2MASS1523+30	&L8				&21.53	&19.18	&15.94	&2.35	&3.24	\\	
SDSS0857+57		&L8			&20.28	&17.86	&14.78	&2.42	&3.08	\\	
SDSS1331-01		&L8			&20.59	&18.66	&15.32	&1.93	&3.34	\\	
2MASS0310+16	&L9				&21.34	&18.89	&15.83	&2.45	&3.06	\\	
2MASS0908+50	&L9				&19.85	&17.48	&14.40	&2.37	&3.08	\\	
2MASS0328+23	&L9.5			&21.78	&19.37	&16.34	&2.41	&3.03	\\		
SDSS0805+48		&L9.5			&19.74	&17.96	&14.67	&1.78	&3.29	\\	
SDSS0830+48		&L9				&20.66	&18.26	&15.21	&2.40	&3.05	\\
SDSS0151+12		&T0.5		&22.49	&19.56	&16.24	&2.93	&3.32	\\		
SDSS0423-04		&T0				&19.73	&17.38	&14.29	&2.35	&3.09	\\
SDSS0837-00		&T1			&23.86	&20.43	&16.89	&3.43	&3.54	\\	
SDSS0758+32		&T2			&21.46	&18.49	&14.77	&2.97	&3.72	\\	
SDSS1254-01		&T2				&21.68	&18.11	&14.65	&3.57	&3.46	\\
SDSS1214+63		&T3.5			&23.20	&20.24	&16.16	&2.96	&4.08	\\	
SDSS1750+17		&T3.5		&22.59	&19.79	&16.13	&2.80	&3.66	\\		
SDSS1021-03		&T3			&22.12	&19.42	&15.87	&2.70	&3.55	\\	
2MASS0559-14	&T4.5			&21.10	&17.63	&13.56	&3.47	&4.07	\\		
SDSS0926+58		&T4.5		&22.16	&19.25	&15.46	&2.91	&3.79	\\		
SDSS1624+00		&T6				&22.41	&19.34	&15.19	&3.07	&4.15	\\
SDSS1346-00		&T6.5			&22.90	&19.54	&15.48	&3.36	&4.06	\\	
2MASS0727+17	&T7				&22.73	&19.47	&15.18	&3.26	&4.29	\\	
Gl570D			&T7.5			&23.20	&19.26	&14.75	&3.94	&4.51	\\	
2MASS0415-09	&T8			&22.97	&19.80	&15.29	&3.17	&4.51	\\		             
\hline                                  
\hline                                  
\end{tabular}
\end{table}
  
Owing to the high dispersion of brown dwarf colours, the classification of one candidate based on 
its photometry is not reliable. However, the colours can be used to classify the object within a large
category such as early L or late L, the overall classification of the objects being
statistically significant. Table~\ref{tab-classification} gives the locus of the spectral classes in the colour-colour diagram.

In the L-dwarf domain, bright L dwarfs discovered in SDSS or 2MASS and with known spectral type that are
also identified on the CFBDS images allow us to validate the colour-based classification.
In the T-dwarf domain, we compared the synthetic colours as a function of spectral type with the 
real colours of T-dwarf candidates that we followed-up spectroscopically. Their locus in the colour-colour diagram is shown by crosses with error bars in Fig.~\ref{iz-zJ_synth}. While the agreement is good for early T-dwarfs, it is poor for the late T-dwarfs where the observed $z'-J$ colours are bluer than the synthetic ones. In particular, some of the coldest known object are located in the same region of the colour-colour diagram as mid-T dwarfs.

%%This is expected as water and methane absorption bands are stronger for late-T where they saturate around the T8 spectral type. Methane bands being important in the near-infrared bands, the resulting near-infrared colours are bluer.

Several reasons can be invoked to explain the discrepancy between observed and synthetic colours of late-T. First, the
number of available spectra of late T-dwarfs is very small (only five
with spectral type later than T6). Because the colour dispersion is quite
high for a given spectral type, the mean synthetic colour can be
biased by one peculiar object.
Next, the synthetic colours of late T-dwarfs may be uncertain, because for these extreme objects with very narrow flux peaks a small error in the optical transmission can lead to large colour uncertainties.
However, as we performed spectroscopic follow-up for all candidates with late-T colours, we used their spectral type for classification instead of colours.

%%more efficient in detecting T-dwarfs with $z'-J$ colours redder than those detected in CFBDS.

\begin{table}[h]
\caption{Definition of spectral classes from the $i'-z'$ and $z'-J$ colours.}             
\label{tab-classification}      
\centering                          
\begin{tabular}{l l l}        
\hline\hline                 
Spectral class &$i'-z'$ range &$z'-J$ range\\    
\hline                        
$<$M8&---&$<2.5$\\
M8-L4.5&$<2.0$&$>2.5$\\
L5-T0&2.0$-$2.6&$>2.5$\\
T0.5-T4.5&$>2.6$&2.5$-$3.8\\
$>$T4.5&---&$>3.8$\\
\hline                                  
\hline                                  
\end{tabular}
\end{table}

\subsection{The sample}
\label{sample}

The CFBDS is composed of several patches -- contiguous areas on the sky -- with sizes ranging from 9 to 79 \sqdeg. The galactic latitude over a patch is nearly constant, as well as the stellar density and the interstellar reddening. The reddening is low for all patches ($\sim0.011\pm0.009$ for RCS2 patches and $\sim0.020\pm0.009$ for CFHTLS patches). %Moreover the different images of the same patch are often obtained the same night. 
Therefore each patch constitutes a sample of homogenous data.

Because the priority was given for the reddest candidates ($i'-z'=2$ or redder) for  $J$-band follow-up, we can now build a complete sample of late L and T dwarf candidates over 16 patches with a total effective area of 444 \sqdeg\ (Table~\ref{fields}). The sample contains all candidates with $z'<22.5$ and $i'-z'>2.0$ detected in this area.

\begin{table}[h]
\caption{Patches used to build a complete sample. All candidates within our selection box ($z'<22.5$ and $i'-z'>2.0$) in these patches have $J$-band photometry.}             
\label{fields}      
\centering                          
\begin{tabular}{c c c c c c c}        
\hline\hline                 
patch & survey & alpha& delta & l  & b &area \\    
 && hh:mm) & (deg) & (deg) & (deg) &(\sqdeg)\\    
\hline                        
   0047 &RCS-2 & 00:47 & $+00$ &120 &-63& 47 \\      
   0310 &RCS-2 & 03:10 & $-16$  &201 &-56  & 56 \\
   0357 &RCS-2 & 03:57 & $-07$   &197 &-42 & 27 \\
   1040 &RCS-2 & 10:55 & $+57$   &149 &+54 & 30 \\ 
   0112 &CFHTLS & 00:45 & $+34$  &121 &-29  & 18 \\ 
   1237 &CFHTLS & 08:30 & $+19$ &206 &30 &9 \\      
   1319 &CFHTLS & 09:15 & $+17$  &213 & 39  & 15 \\
   1512 &CFHTLS & 10:30 & $+11$  &232 &53  & 14 \\
   1514 &RCS-2 & 15:14 & $+06$  &7 &50 & 48 \\
   1645 &RCS-2 & 16:45 & $+40$   &64 &41 & 28 \\ 
   2017 &CFHTLS & 13:35 & $-08$  &231 &53  & 13 \\
   2143 &RCS-2 & 21:43 & $-01$   &55 &-38 & 79 \\
   2265 &CFHTLS & 15:00 & $-17$ &342 &36  & 14 \\
   3071 &CFHTLS & 19:30 & $-22$   &17 &-18 & 13 \\ 
   3303 &CFHTLS & 22:10 & $-18$   &38 &-52 & 14 \\
   3400 &CFHTLS & 22:30 & $+03$  &69 &-44  & 19 \\ 
\hline                                  
\hline                                  
\end{tabular}
\end{table}

\begin{figure}[!h]
\begin{center}
\includegraphics[scale=0.45,clip=]{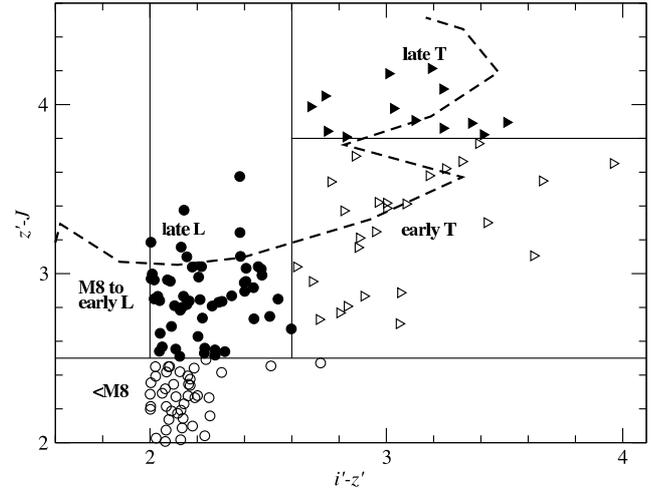}
\caption{
$z'-J$ versus $i'-z'$ diagram of the brown dwarf candidates in our sample with complete $J$-band follow-up. Open circles: dwarfs earlier than M8. Filled circles: M8 to L dwarfs. Open triangles: early T dwarfs. Filled triangles: late T dwarfs. The dotted line shows the colour-colour relation derived in 
 Sect.~\ref{classification}. The solid lines show the different spectral type regions defined in Table~\ref{tab-classification}. Note that to build the T dwarfs luminosity-function we did not use
this colour-based classification but used the spectral type obtained with spectroscopic
follow-up whenever available.
\label{iz-zJ_sample}
}
\end{center}
\end{figure}

This sample contains 249 objects that are candidates cooler than L5 on the basis of their $i'-z'$ colour. However, due to the photometric errors, many contaminants with true colour $i'-z'<2.0$ enter this sample, such as the more numerous M and early L dwarfs. A classification of the candidates is made on the basis of their position in the $i'-z'$/$z'-J$ diagram, as shown in Fig.~\ref{iz-zJ_sample} and explained in Sect.~\ref{classification} (see Table~\ref{tab-classification} for a summary).
Based upon this classification, 102 of the 249 objects remain L5 and later dwarf candidates (Table~3), the others are dwarfs earlier than M8, artefacts, or quasars. Dwarfs of spectral type $>$ M8 to L5
are theoritically not retained by the $i'-z'>2.0$ criteria. But
they could statistically contaminate our $>$L5 sample if their $i'-z'$
colour is reddened by photometric errors; then the $z'-J$ criteria
will not reject them because they have similar colours as the $>$L5 dwarfs in this wavelength
bandpass. This contamination has to be carefully estimated (Sect.~\ref{contaminants}). The distribution of objects within the spectral classes is given in Table~\ref{classif-sample}.

\begin{table*}[h]
\caption{Photometry of L5 and later dwarf candidates in the MegaCam photometric system. For objects without detection in the $i'$-band, a lower limit is given for the $i'-z'$ colour. The spectral type is given for dwarfs with spectroscopic follow-up.}             
\label{tabsample}      
\centering                          
\begin{tabular}{c c c c c c c c c c c c}        
\hline\hline                 
$\alpha$ J2000	&$\delta$ J2000	&$z'$	&$z'_{err}$	&$i'$	&$i'_{err}$	&$i'-z'$	&$i'-z'_{err}$	&$J$	&$J_{err}$	&$z'-J$	&Spectral class \\
\hline
00:18:51.37	&+34:27:25.67	&22.051	&0.086	&25.477	&0.149	&3.426		&0.172	&18.666	&0.046	&3.385  &early T       \\
00:32:14.37	&+02:56:04.66	&21.875	&0.052	&24.019	&0.092	&2.144		&0.105	&18.500	&0.021	&3.375	&late L        \\
00:34:02.82	&-00:52:05.74	&22.114	&0.065	&26.373	&0.561	&$>$3.030	&0.565	&18.137	&0.012	&3.977	&T9/Y$^a$  \\
00:34:48.76	&-02:05:00.62	&22.141	&0.069	&24.431	&0.165	&2.290		&0.179	&19.311	&0.030	&2.830	&late L        \\
00:40:37.76	&+03:53:16.67	&21.313	&0.023	&23.492	&0.052	&2.179		&0.057	&18.274	&0.019	&3.039	&late L        \\
00:41:48.71	&-01:33:53.81	&22.230	&0.081	&24.316	&0.097	&2.086		&0.126	&19.275	&0.046	&2.955	&late L        \\
00:45:26.33	&+35:47:11.44	&21.069	&0.039	&23.785	&0.030	&2.716		&0.049	&18.340	&0.053	&2.729	&early T       \\
00:49:28.38	&+04:40:58.88	&18.675	&0.004	&22.297	&0.015	&3.623		&0.016	&15.569	&0.004	&3.106  &early T       \\
00:51:07.95	&+01:41:05.98	&21.945	&0.058	&24.529	&0.140	&2.584		&0.152	&18.665	&0.023	&3.280	&late L        \\
00:53:04.70	&-02:00:25.30	&22.069	&0.079	&24.073	&0.073	&2.004		&0.108	&18.884	&0.051	&3.185	&late L        \\
00:53:36.13	&+04:52:14.40	&22.361	&0.080	&99.000	&0.575	&2.230		&0.581	&19.576	&0.049	&2.786	&late L        \\
00:53:50.41	&+05:00:02.15	&22.485	&0.085	&24.525	&0.579	&$>$2.040	&0.586	&19.944	&0.059	&2.541	&late L        \\
00:57:00.41	&-03:29:43.57	&19.783	&0.010	&22.586	&0.019	&2.803	        &0.022	&16.899	&0.014	&2.914	&early T       \\
00:59:10.83	&-01:14:01.31	&21.935	&0.052	&28.050	&0.549	&$>$3.240	&0.552	&18.075	&0.013	&3.860	&T9/Y$^b$      \\
01:04:07.71	&-00:53:27.62	&19.427	&0.006	&21.934	&0.012	&2.507		&0.014	&16.680	&0.003	&2.747  &late L	     \\
01:06:08.60	&+34:05:20.24	&22.007	&0.010	&24.323	&0.048	&2.317		&0.111	&19.468	&0.028	&2.539	&late L        \\
01:11:59.09	&+33:08:07.95	&21.195	&0.046	&23.602	&0.032	&2.407		&0.056	&18.163	&0.010	&3.032	&late L        \\
02:52:32.48	&-16:22:55.40	&22.149	&0.076	&24.305	&0.161	&2.156		&0.179	&19.332	&0.033	&2.817	&late L        \\
02:52:42.95	&-17:32:25.30	&21.413	&0.036	&23.871	&0.130	&2.458		&0.134	&18.372	&0.031	&3.041	&late L        \\
02:54:01.67	&-18:25:29.25	&22.184	&0.090	&99.000	&0.583	&$>$2.680	&0.590	&18.197	&0.013	&3.987	&T6.5            \\
02:54:18.26	&-13:52:37.15	&21.599	&0.051	&23.904	&0.089	&2.305		&0.103	&18.763	&0.026	&2.836	&late L        \\
02:54:20.68	&-17:07:44.39	&21.630	&0.040	&24.031	&0.151	&2.401		&0.156	&18.735	&0.026	&2.895	&late L        \\
02:54:45.64	&-14:03:14.86	&20.624	&0.021	&22.887	&0.039	&2.263		&0.045	&17.816	&0.015	&2.808	&late L        \\
02:55:58.46	&-17:30:20.29	&21.021	&0.027 	&23.907	&0.138	&2.886		&0.141	&17.810	&0.028	&3.211	&T1       \\
02:56:12.28	&-12:57:21.39	&20.588	&0.024	&22.806	&0.046	&2.218		&0.052	&17.547	&0.008	&3.041	&late L        \\
02:57:18.08	&-12:48:53.13	&21.439	&0.044	&24.404	&0.145	&2.965		&0.152	&18.018	&0.016	&3.421	&T3.5       \\
02:58:05.92	&-14:55:34.34	&21.229	&0.025	&25.189	&0.524	&$>$3.960	&0.525	&17.578	&0.033	&3.651	&T1.5          \\
02:58:40.60	&-18:26:48.30	&20.960	&0.031	&23.827	&0.124	&2.868		&0.128	&17.339	&0.015	&3.621	&T3       \\
02:59:47.10	&-11:36:19.11	&22.179	&0.080	&24.416	&0.102	&2.237		&0.130	&19.436	&0.085	&2.743	&late L        \\
03:01:30.53	&-10:45:04.38	&21.268	&0.032	&24.883	&0.531	&$>$3.360	&0.532	&17.379	&0.007	&3.889	&T6            \\
03:01:48.80	&-13:49:48.91	&20.759	&0.030	&23.196	&0.044	&2.438	        &0.054	&17.557	&0.014	&3.202	&late L        \\
03:02:25.88	&-14:41:25.47	&21.704	&0.064	&28.240	&0.561	&$>$3.190	&0.564	&17.491	&0.010	&4.213	&T5.5            \\
03:02:26.67	&-14:37:19.23	&21.264	&0.043	&24.674	&0.175	&3.410		&0.180	&17.442	&0.009	&3.822	&T4            \\
03:04:21.58	&-13:58:36.75	&20.935	&0.032	&23.091	&0.041	&2.156		&0.052	&17.836	&0.020	&3.099	&late L        \\
03:07:19.75	&-17:15:32.25	&21.702	&0.058	&23.921	&0.099	&2.219		&0.115	&18.637	&0.035	&3.065	&late L        \\
03:10:11.40	&-16:22:25.32	&22.207	&0.062	&24.349	&0.160	&2.142		&0.172	&19.341	&0.074	&2.866	&late L        \\
03:10:28.48	&-10:29:28.95	&21.394	&0.048	&23.676	&0.084	&2.282		&0.096	&18.726	&0.036	&2.668	&late L        \\
03:18:59.69	&-17:05:27.75	&21.282	&0.036	&23.489	&0.092	&2.206		&0.099	&18.052	&0.021	&3.230	&late L        \\
03:19:00.22	&-17:10:36.03	&21.266	&0.038	&24.219	&0.180	&2.954		&0.184	&17.985	&0.022	&3.281	&early T       \\
03:24:38.38	&-12:40:30.93	&22.155	&0.059	&24.319	&0.145	&2.165		&0.156	&19.400	&0.049	&2.754	&late L        \\
03:26:02.49	&-13:39:27.71	&19.443	&0.011	&21.590	&0.013	&2.148		&0.017	&16.772	&0.005	&2.671	&late L        \\
04:00:27.15	&-08:57:53.99	&21.387	&0.038	&23.473	&0.064	&2.086		&0.074	&18.593	&0.042	&2.794	&late L        \\
04:04:20.89	&-08:18:29.14	&21.458	&0.046	&23.476	&0.053	&2.018		&0.071	&18.495	&0.021	&2.963	&late L        \\
08:30:09.57	&+19:04:27.99	&21.622	&0.038	&23.713	&0.081	&2.091		&0.090	&18.935	&0.021	&2.687	&late L        \\
08:58:33.25	&+17:34:53.47	&22.371	&0.056	&24.716	&0.172	&2.346		&0.181	&19.502	&0.037	&2.869  &late L        \\
09:01:39.82	&+17:40:51.35	&21.566	&0.028	&24.331	&0.128	&2.765		&0.131	&18.023	&0.010	&3.543	&T4       \\
09:04:49.60	&+16:53:47.08	&21.679	&0.032	&25.336	&0.531	&$>$3.250	&0.532	&18.058	&0.011	&3.621  &T4            \\
09:22:50.12	&+15:27:41.44	&22.384	&0.055	&25.907	&0.552	&$>$2.180	&0.555	&18.810	&0.020	&3.574	&T7        \\
10:28:41.01	&+56:54:01.91	&22.246	&0.077	&29.159	&0.572	&$>$2.740	&0.577	&18.195	&0.011	&4.051	&T8            \\
10:30:18.74	&+09:41:44.72	&21.662	&0.038	&24.349	&0.110	&2.687		&0.116	&18.709	&0.018	&2.953	&early T       \\
10:39:44.90	&+10:07:37.46	&21.448	&0.030	&24.502	&0.132	&3.054		&0.135	&18.744	&0.021	&2.704	&early T       \\
10:40:55.63	&+08:40:05.38	&21.840	&0.044	&24.071	&0.087	&2.232		&0.097	&19.280	&0.026	&2.560	&late L        \\
10:42:09.98	&+58:08:56.63	&21.843	&0.045	&26.328	&0.543	&$>$3.010	&0.545	&17.660	&0.007	&4.183	&T6.5            \\
10:55:03.16	&+58:08:27.74	&22.376	&0.094	&24.419	&0.154	&2.043		&0.180	&19.729	&0.036	&2.647	&late L        \\
13:24:00.09	&-8:11:23.32	&22.123	&0.062	&24.666	&0.179	&2.542		&0.190	&19.274	&0.032	&2.849	&late L        \\
13:29:45.70	&-8:27:30.82	&22.447	&0.076  &24.659	&0.168	&2.212		&0.184	&19.601	&0.075	&2.846	&late L        \\
13:32:40.95	&-8:23:14.34	&22.343	&0.070	&24.546	&0.146	&2.203		&0.162	&19.715	&0.040	&2.628	&late L        \\
13:34:07.01	&-9:23:39.79	&22.446	&0.080	&24.572	&0.152	&2.126		&0.172	&19.935	&0.053	&2.511	&late L        \\
14:50:44.96	&+09:21:08.72	&22.166	&0.070	&25.225	&0.566	&$>$2.820	&0.570	&18.795	&0.032	&3.371	&T3.5            \\
14:58:47.93	&+06:14:02.9	&21.553	&0.061  &23.960	&0.113  &2.383		&0.120	&18.215	&0.027	&3.338  &T2            \\
14:59:35.30	&+08:57:51.57	&21.693	&0.056	&24.522	&0.157	&2.829		&0.167	&17.885	&0.015	&3.808	&T4.5          \\
15:00:00.50	&-18:24:07.35	&21.626	&0.045	&26.360	&0.543	&$>$3.000	&0.545	&18.209	&0.017	&3.417  &T4.5          \\
15:02:10.19	&+03:50:55.51	&22.086	&0.081	&25.585	&0.575	&$>$3.060	&0.581	&19.199	&0.099	&2.887	&early T       \\
15:04:11.71	&+10:27:17.9	&20.609	&0.019  &25.980 &0.654  &$>$4.360	&0.650	&16.497	&0.011	&4.112  &T7$^c$            \\
15:04:35.21	&+10:26:38.1	&21.584	&0.042  &24.004 &0.108  &2.399		&0.110	&18.402	&0.038	&3.182  &late L        \\
15:07:24.48	&+07:29:19.11	&20.715	&0.026	&22.748	&0.033	&2.033		&0.042	&17.849	&0.015	&2.866	&late L        \\
15:10:14.33	&+11:13:25.32	&21.015	&0.036  &23.920	&0.063	&2.905		&0.072	&18.148	&0.079	&2.867	&early T       \\
\hline                                  
\hline                                  
\end{tabular}

$^a$ \cite{Warren.2007}. $^b$ \cite{Delorme.2008a}. $^c$ \cite{Chiu.2006}.
\end{table*}

\begin{table*}[h]
\label{tabsample}      
\centering                          
\begin{tabular}{c c c c c c c c c c c c}        
\multicolumn{12}{p{17.8cm}}{{\bf \small \hspace*{-.35cm} Table 3.} \small Continued\vspace*{0.3cm}}\\
\hline\hline                 
$\alpha$ J2000	&$\delta$ J2000	&$z'$	&$z'_{err}$	&$i'$	&$i'_{err}$	&$i'-z'$	&$i'-z'_{err}$	&$J$	&$J_{err}$	&$z'-J$	&Spectral class \\
\hline
15:11:14.59	&+06:07:42.32	&19.199	&0.007	&21.174	&0.011	&1.980		&0.012	&16.016	&0.079	&3.160  &T2$^c$            \\
15:13:24.98	&+09:53:45.05	&20.787	&0.025	&23.261	&0.040	&2.474		&0.047	&17.796	&0.014	&2.991	&late L        \\
15:15:46.59	&+06:17:38.7	&21.152	&0.043  &23.227 &0.064  &2.052		&0.070	&18.346	&0.037	&2.806  &late L        \\
15:18:03.64	&+07:16:46.0	&21.627	&0.055  &24.831 &0.259  &3.181		&0.260	&18.048	&0.027	&3.579  &T2.5          \\
15:19:29.92	&+10:50:59.17	&21.746	&0.073	&23.752	&0.055	&2.005		&0.091	&18.776	&0.029	&2.970	&late L        \\
15:25:14.82	&+11:18:33.16	&21.642	&0.068	&25.065	&0.564	&$>$3.320	&0.568	&17.979	&0.029	&3.663	&T2.5          \\
15:26:55.80	&+03:45:36.2	&21.595	&0.069  &24.750 &0.230  &3.121		&0.200	&17.689	&0.021	&3.906  &T4            \\
15:10:10.21	&+07:54:52.9	&21.482	&0.062  &24.108 &0.140  &2.598		&0.130	&18.575	&0.037	&2.907  &late L        \\
16:36:58.93	&+39:31:51.28	&22.338	&0.088  &27.987	&0.581	&$>$2.880	&0.588	&19.184	&0.028	&3.154	&early T       \\
16:39:55.89	&+38:55:19.17	&22.296	&0.078	&25.130	&0.200	&2.834		&0.215	&19.489	&0.032	&2.810	&early T       \\
16:43:57.37	&+41:57:40.62	&22.388	&0.098	&24.859	&0.184	&2.471		&0.209	&19.361	&0.023	&3.027  &late L        \\
16:58:43.55	&+38:11:55.73	&22.151	&0.069	&24.532	&0.100	&2.381		&0.121	&18.908	&0.015	&3.243	&late L        \\
20:37:37.07	&-19:22:02.90	&21.860	&0.050	&24.490	&0.210	&2.620		&0.220	&18.500	&0.039	&3.360	&T0          \\
20:38:41.41	&-18:50:12.31	&21.983	&0.072	&25.066	&0.344	&3.083		&0.351	&18.581	&0.031	&3.402	&T3          \\
20:48:03.61	&-18:32:12.79	&20.303	&0.017	&23.812	&0.097	&3.509		&0.099	&16.442	&0.009	&3.862	&T4.5            \\
21:22:43.69	&+04:29:41.98	&21.164	&0.028	&24.165	&0.158  &3.001		&0.160	&17.777	&0.016	&3.387	&T2          \\
21:24:09.76	&-00:04:52.93	&22.095	&0.070	&24.317	&0.129	&2.222		&0.147	&19.358	&0.066	&2.737	&late L        \\
21:24:13.96	&+01:00:02.40	&19.907	&0.011	&23.146	&0.077	&3.239		&0.077	&15.815	&0.004	&4.092	&T5$^d$            \\
21:27:02.19	&+00:23:44.70	&21.279	&0.039  &25.252	&0.537	&3.660		&0.539	&18.127	&0.015	&3.152	&T3.5            \\
21:32:21.96	&-00:08:59.88	&21.182	&0.022	&23.287	&0.051	&2.104		&0.055	&18.372	&0.034	&2.810	&late L        \\
21:32:59.92	&-01:38:27.17	&22.209	&0.068	&24.485	&0.193	&2.276		&0.205	&19.691	&0.085	&2.518	&late L        \\
21:36:07.09	&-02:22:32.66	&20.910	&0.032	&23.077	&0.062	&2.166		&0.070	&18.072	&0.025	&2.838	&late L        \\
21:39:26.90	&+02:20:23.58	&18.440	&0.004	&21.450	&0.046	&2.750		&0.046	&14.710	&0.003	&3.730	&early T       \\
21:40:48.01	&+00:21:35.54	&21.794	&0.047	&23.923	&0.126	&2.128		&0.134	&19.011	&0.032	&2.783	&late L        \\
21:41:24.58	&+03:22:49.34	&21.854	&0.057	&23.963	&0.173	&2.109		&0.183	&19.300	&0.034	&2.554  &late L        \\
21:41:27.99	&-00:28:41.82	&22.210	&0.080	&24.284	&0.159	&2.074		&0.178	&19.247	&0.021	&2.963	&late L        \\
21:41:39.77	&-03:37:39.07	&21.889	&0.064	&24.492	&0.560	&$>$2.240	&0.564	&18.724	&0.017	&3.140  &T1            \\
21:47:20.95	&-00:55:33.02	&21.998	&0.053	&24.130	&0.148	&2.132	        &0.157	&18.841	&0.061	&3.157	&late L        \\
21:52:30.47	&-00:45:04.70	&22.363	&0.073	&24.763	&0.568	&$>$2.010	&0.573	&19.365	&0.032	&2.998	&late L        \\
21:53:52.15	&-01:38:55.38	&22.093	&0.090	&24.135	&0.182	&2.041		&0.203	&19.253	&0.029	&2.840	&late L        \\
21:54:54.06	&-01:44:16.38	&21.454	&0.055	&23.860	&0.142	&2.406		&0.153	&18.500	&0.026	&2.954	&late L        \\
21:59:18.86	&+03:05:07.06	&19.844	&0.014	&23.145	&0.081	&$>$3.300	&0.082	&17.300	&0.008	&2.544	&early T       \\
22:12:45.12	&-12:30:38.68	&22.209	&0.063	&24.484	&0.188	&2.275		&0.199	&19.683	&0.042	&2.526	&late L        \\
22:17:06.13	&-12:05:41.13	&21.352	&0.026	&23.370	&0.065	&2.018		&0.070	&18.501	&0.019	&2.851	&late L        \\
22:22:45.34	&-11:07:31.62	&20.968	&0.024	&23.105	&0.067	&2.136		&0.071	&18.168	&0.021	&2.800	&late L        \\
22:38:56.30	&+03:49:47.00	&21.562	&0.033	&25.141	&0.532	&3.390		&0.533	&17.804	&0.020	&3.759	&T2            \\
\hline                                  
\hline                                  
\end{tabular}

$^c$ \cite{Chiu.2006}. $^d$ \cite{Knapp.2004}.
\end{table*}

\begin{table}[h]
\caption{Preliminary classification of the sample based on the $i'-z'$ and $z'-J$ colours. The spectral class definitions are given in Table~\ref{tab-classification}. Note that we expect most contamination to occur in the late L class from the M8 to early L class. The limiting distance of detection is also given.}             
\label{classif-sample}      
\centering                          
\begin{tabular}{l l l l}        
\hline\hline                 
Object type & number & percentage & distance\\    
\hline  
late T dwarfs &13&5\%&60 pc\\                      
early T dwarfs &28 &11\%&100 pc\\
late L dwarfs &61 &25\%&200 pc\\
$<$M8 dwarfs &77 &31\%&500 pc\\
%quasars &8 &3\%\\
artefacts and quasars candidates &70 &28\%\\
\textbf{total} &\textbf{249}&\\
\hline                                  
\hline                                  
\end{tabular}
\end{table}

\subsection{Completeness}
\label{complete}

Owing to photometric errors, some objects with true colours within our
selection criteria are spread out of our selection box and are not
identified. Given the depth of CFBDS, one cannot estimate the
completeness with a reference sample of objects detected in the
images with previously known magnitude and colours. To obtain a reference sample we built a point spread function (PSF) model for each science image
using \textit{PSfex} (E. Bertin, private communication)
and created fake stars %of a chosen array of magnitudes and colours
with $1.2<i'-z'<4$ and $20<z'<23.5$
with \textit{Skymaker} \citep{Bertin.2009b}. These artificial
stars are thus a PSF model scaled to the desired magnitude, affected
with a realistic photon noise. This model is oversampled so that it
handles correctly the randomised non-integer pixel position of the
fake stars. We then added these stars to the science images from which
the PSF model was derived, which means that any analysis of these
fake stars is affected by the same background noise, bad pixels,
saturated stars, and
cosmic ray distribution as the actual stars are.

To avoid any
significant increase of the image crowding caused by this addition, we
injected a number of artificial stars that is lower than 5\% of the
number of astrophysical sources on the images. This left us with a
statistically comfortable sample of  1~500~000 artificial ultracool
dwarfs. Drawing upon this sample we were able to derive CFBDS's
completeness.

The resulting images, containing both astrophysical sources and fake
stars, went through the same analysis and selection pipelines used to select true
candidates. The injection of fake objects in the science images allows
us to take into account all effects such as cosmic rays, bad pixels,
resampling noise, and possible selection biases when going
through the selection pipeline. The effect of faint
sources being lost due to the presence of a nearby bright star is
also perfectly taken into account without the use of the somewhat arbitrary
masking of a given area around saturated stars: their recovery rate
directly depends on their magnitude and separation to the saturated stars
actually present in  the science images. The injection
of stars into the actual science images is a direct and extremely
robust measure of
completeness regardless of what actually causes the loss of completeness.

\begin{figure}[!h]
\begin{center}
\includegraphics[scale=0.45,angle=0]{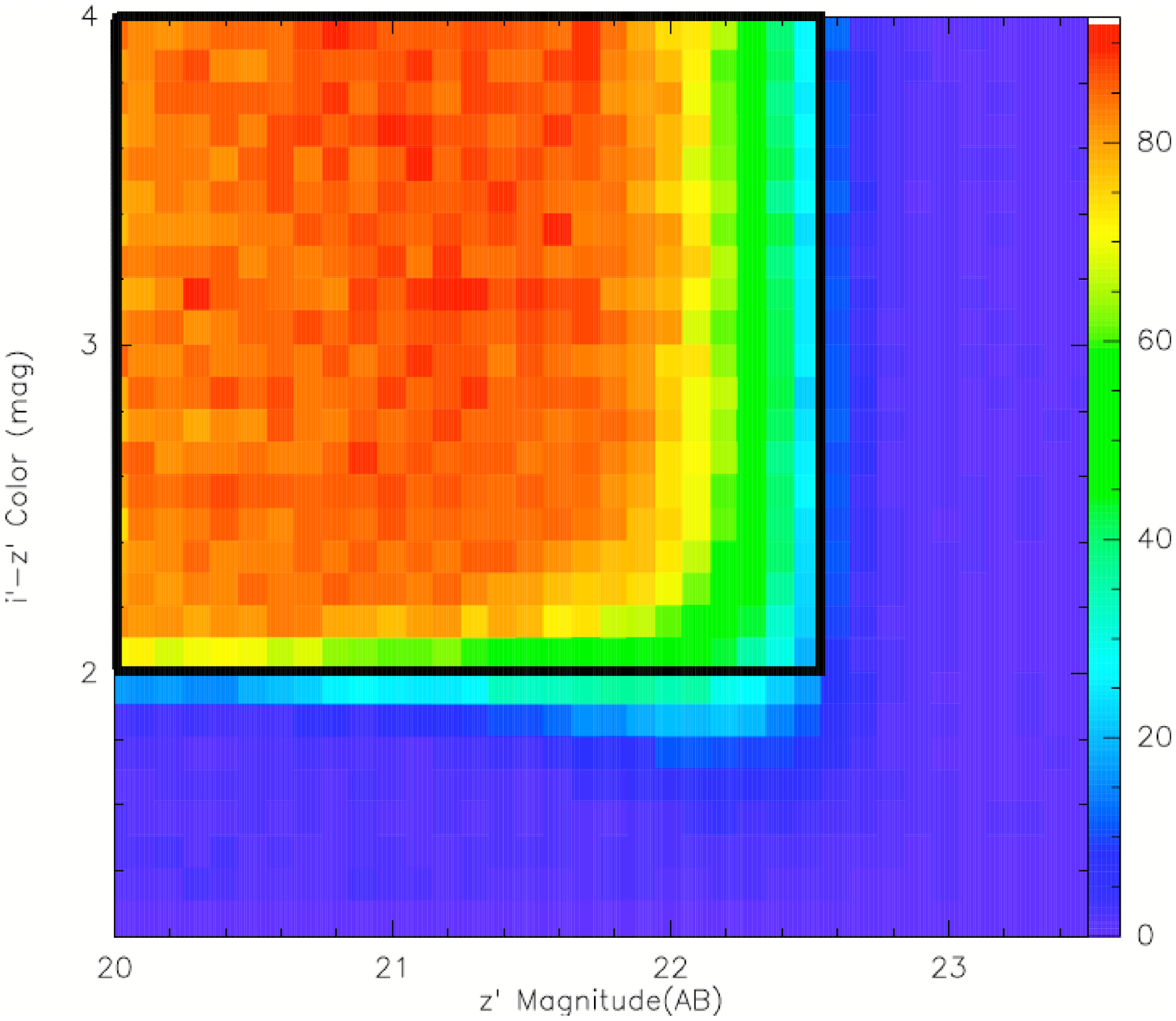}
\caption{
Completeness of the RCS2 component of the CFBDS survey as a function of $z'$ magnitude and $i'-z'$ colour. The colour bar ranges from 0\% (blue, all objects went out of the selection limits) to 100\% (red, all objects detected). The volume-weighted average completeness over our selection criterion (black rectangle, $i'-z'>2.0$ and $z'<22.5$) is 73\%. This plot also highlights the contamination of our 
sample by objects whose actual colours and magnitudes lay outside our selection box but 
are scattered inside by photometric errors.
\label{completeness}
}
\end{center}
\end{figure}

Thus we can count the number of detected objects as a function of
magnitude on the detection image ($z'$-band) and colour. This is done separately
for each patch of the CFBDS. The completeness is given by the fraction of recovered fake
objects at the end of the analysis process. Note that the measured
colours and magnitudes of fake objects are different from the injected
colours because of noise and measurement errors.
Figure~\ref{completeness} shows the completeness averaged over the patches
of the RCS-2 survey as a function of "true" (that is before going through the analysis process) magnitude and  colour. We note
$c_{ij}$ the completeness in each bin $(i,j)$ of 0.1 mag in magnitude
and colour. Note that this completeness does not merely reflect the
number of objects detected, but the number of detected objects which
went through our whole pipeline and were selected as ultracool dwarfs
candidates with a star-like shape, a signal-to-noise ratio over 10, a
measured $i'-z'$ colour over our selection criterion ($i'-z'>2.0$ for the sub-sample studied here),
%(or 2.0, depending on the sample),
and a measured $z'$ magnitude below 22.5.

\begin{figure}[!h]
\begin{center}
\includegraphics[scale=0.5,angle=270,clip=]{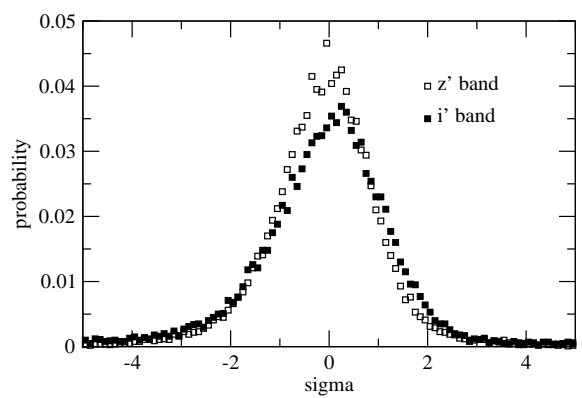}
\caption{
Probability error distribution for one of the RCS2 patches, in the $i'$ and $z'$ bands. The errors are given in number of $\sigma$. They are computed by comparing the magnitudes of the fake cool dwarfs when injected in the science images with their measured magnitudes.
\label{error}
}
\end{center}
\end{figure}

Moreover, the measured magnitudes are compared with the "true" ones to derive an
error probability distribution (Fig.\ref{error}). It gives the probability for one object in a given patch to have a different measured photometry by a given number of $\sigma$ from its true photometry. The result is very close to a Gaussian distribution, the only difference is found in the wings  at more than 5 $\sigma$, where the number of deviations stands at 1-2\%, several orders of magnitude above a Gaussian distribution. 
This is due to blended objects with galaxies, bad pixels, etc.
This distribution, which also takes into account the correlation between pixels caused when resampling any images, will be used when computing the contamination of the sample (Sect.~\ref{contaminants}).

\subsection{Contaminants}
\label{contaminants}
Reddened stars are not possible contaminants because the CFBDS fields are in a low galactic extinction area ($i' - z'$ galactic reddening $<$ 0.02 mag). Giant stars are not expected in our sample because with the apparent
magnitude of our sample ($z'$ between 18.5 and 22.5) they would be
located at 100 kpc to 1 Mpc, well outside of the Milky Way. Actually, we noticed a contamination by giant stars for fields
close to Messier galaxies that we had to eliminate from the survey.
 
Extremely reddened galaxies could potentially have $i'-z'$ and $z'-J$ colours the same as 
those of brown dwarfs. These galaxies would have to be at a fairly high redshift ($z>1$) 
for two reasons: (i) higher redshift means shorter rest-frame wavelengths and therefore a 
lower $A_V$ to give the red observed colours; (ii) the CFBDS objects are unresolved in 
good seeing (typically $0.6-0.8"$) and lower redshift galaxies would be easily resolved. 
However, the high redshift necessary, coupled with the fairly bright $J$ magnitudes of 
the CFBDS objects and the fact that the red $z'-J$ indicates significant extinction still 
at $J$ (e.g. about 3 magnitudes of extinction for rest-frame $E(B-V)=1$),  means these 
galaxies would need to have extremely high intrinsic absolute magnitudes and stellar 
masses (%$>10\,M_*$ or 
$>2\times 10^{12}\,M_\odot$). Galaxies of this mass are extremely 
rare at redshift $z\sim 1$ \citep{Drory.2009}, and are composed of evolved stellar 
populations rather than dusty starbursts \citep{Taylor.2009}. These massive galaxies would likely be 
resolved by ground-based seeing out to redshift $z \sim 2$ \citep{Mancini.2010}. In conclusion, we find 
that high-redshift unresolved galaxies cannot appear as red as the L and T dwarfs and do 
not cause a significant contaminant.

As already mentioned, high-redshift quasars have the same $i'-z'$ colour as brown dwarfs but are easily
removed form the sample through their bluer $z'-J$ colour by about one magnitude. Moreover, they are rare objects, 
so their contamination is indeed negligible.

As shown in the colour-colour graph in Fig.~ \ref{iz-zJ_synth} our sample selects L5 and
cooler dwarfs. However, each Megacam field shows on average 50000 sources,  
of which about one is selected as a brown dwarf candidate, meaning that
non-brown dwarf sources outnumber brown dwarfs by more than a factor
10000, making 3 or 4 $\sigma$ photometric noise scattering a probable
source of contamination, particularly in
the $i'$ band where brown dwarfs are faint and observed at low
signal-to-noise ratio and a fraction of earlier type dwarfs with true
colour $i'-z' < 2.0$ are sent within the $i'-z'>2.0$ sample (see Fig.~\ref{completeness}). This
contamination has to be estimated carefully because earlier dwarfs are much
more numerous in a limited-magnitude sample than the later ones. 

There is only a small probability for a given object to have large photometric deviations in two independent filters. Therefore the $J$-band observations at a good S/N ratio are useful to identify a significant fraction of the contaminants. 
All objects with $z'-J < 2.5$
are removed from the sample. This selection rejects (1) artefacts with
obviously
no J-band detection, (2) quasars, and (3) dwarfs with true $i'-z'$ colour
lower than 1.3, corresponding to spectral type earlier than M8, and
reddening by photometric noise.

The contamination in $z'-J$ does exist but has a
much lower impact on our study for several reasons: (i) we only obtain $J$ magnitudes for
the selected candidates, 249
objects, which make several sigma noise excursions
non-significant. This is especially true because the signal-to-noise ratio 
in $J$ is much better than in $i'$, leading to much smaller
photometric errors, typically in the 0.02-0.1 magnitude range. 
(ii) Because
the colour spectral type relation is very steep at the M7-M8.5
transition, going from $z'-J$=2.3 to $z'-J$=2.9, we are confident that
possible contamination from earlier than M8 dwarfs part of our 249-strong
brown dwarf candidates sample into our $z'-J$
confirmed brown dwarf sample trough 4 to 8 $\sigma$ noise excursion is
marginal. (iii) Our sample is $z'$ selected, so the flux limit bias or Eddington bias that 
occurs due to a scattering of sources fainter than our $z'$ magnitude limit into the sample will 
act to increase the number of high $i'-z'$ objects, but decrease the number of high $z'-J'$ 
objects. In practice this bias is very small in both directions because we have a 
10 $\sigma$ $z'$ limit. (iv) The main reason the scattering from low $i'-z'$ to high $i'-z'$ is so bad 
is because the number density of objects with true low $i'-z'$ is so much higher than 
those with true high $i'-z'$. The opposite is actually true for the L and T dwarfs: in our $i'-z'$ selected
sample, there are 
more objects with true high $z'-J$ (L and T dwarfs) than low $z'-J$ (z$\sim$6 quasars).

%\begin{table}[h]
%\caption{Number of objects as a function of colour and magnitude computed from the luminosity function of \cite{cruz.2007}.}             
%\label{sample1.3-1.7}      
%\centering                          
%\begin{tabular}{l l l l l}        
%\hline\hline                 
%$i'-z'$ &spectral type &$M_J$  &$Mz'$	&number density\\    
%	 &			&		&		&(10$^{-3}$ stars pc$^{-3}$ mag$^{-1}$)\\
%\hline                        
%1.3-1.5 &M8.5-L1 &11-11.5 &14.1	&1.7\\
%1.5-1.6 &L1-L3 	    &11.5-13 &15.0	&2.5\\
%1.6-1.7 &L3-L4.5 &13-13.5  &16.4	&0.6\\
%\hline                                  
%\hline                                  
%\end{tabular}
%\end{table}

The $z'-J$ cut cannot distinguish ultracool dwarfs with true colour $1.3 < i'-z' < 2.0$ 
because they have the same $z'-J$ colour
as late L and early T dwarfs. We therefore need
to evaluate the number of potential contaminants that populate this colour range and can leak into our 
selection through photometric errors.
The estimate of the number of contaminants with true colour
$1.7<i'-z'<2.0$ is done in two steps as explained below and is done 
separately for each patch.

\begin{enumerate}
%\item We neglect the contamination by objects with colour
%$1.3<i'-z'<1.7$. Indeed, although their colours are closer to our
%selection limit, they are much less numerous than the earlier
%ones. To check that the contamination is negligible, we extracted all
%sources with $1.3<i'-z'<1.7$ in each patch down to $z'=23$. Thus we
%draw randomly a photometric error in the $z'$ and $i'$ bands
%following the probability error distributions obtained from the fake
%ultracool dwarfs (Fig.\ref{error} and \S~\ref{complete}). We apply
%the errors to the detected objects. It appears that the objects with
%resulting colour and magnitude within our selection box ($i'-z'>2.0$
%and $z'<$22.5) are very few.

\item Six patches of the CFBDS have $J$-band follow-up for all the
  candidates with measured colour $i'-z'>1.7$. These patches
  contain 228 candidates over a total area of 199\sqdeg. The $z'-J$ diagnostic shows that 35\% 
  of these objects have indeed spectral type later than M8; the 65\% another objects
are $<$M8, artefacts or quasars. We assume that this value of 35\% is
representative of the fraction of $>$M8 in the $1.7<i'-z'<2.0$ sample 
for all the patches. These $>$M8 dwarfs are the main sources of contaminations in
our $>$L5 stellar count

\item Thus we compute the probability to get a M8 to L4.5 dwarf with true colour $1.7<i'-z'<2.0$ but resulting colour $i'-z'>2.0$.
We draw a photometric error in the $i'$ and $z'$ bands for  35\% of the objects with $1.7<i'-z'<2.0$. Objects with resulting $i'-z'>2.0$ and $z'<22.5$ are the contaminants. In practical, we added a photometric
  error to all objects with $1.7<i'-z'<2.0$, and each object that enters our selection box
  will count for only 0.35 contaminant, which provides a smoother
contaminant distribution.

\end{enumerate}

%In this process, the contamination by objects with true colour $1.3<i'-z'<1.7$ is taken into account in an indirect way. Indeed, the sample with $1.7<i'-z'<2.0$ is considered as a sample with true colour $1.7<i'-z'<2.0$.  However, this sample is itself contaminated by objects with $1.3<i'-z'<1.7$. Hence adding an error to the $1.7<i'-z'<2.0$ sample makes objects with true colour $1.3<i'-z'<1.7$ enter into the $i'-z'>2.0$ sample. We thus overestimate the number of potential contaminants in the $1.7<i'-z'<2.0$ range while underestimating the contamination by not taking into account potential contaminants in the  $1.3<i'-z'<1.7$ range. That these two approximations are qualitatively low and tend to compensate each other is relatively straightforward. 
This analysis underestimates the number of contaminants because we do not 
consider direct contamination from the $1.3<i'-z'<1.7$  colour range to $i'-z'>2.0$, but also  
overestimates the number of contaminants because some objects within our $1.7<'i-z'<2.0$ contaminants sample are 
themselves contaminants with bluer true colours. However these effects 
%on the contamination of contaminants 
are not a significant problem, especially because they tend 
to compensate each other: 
%However the fake objects allow for the numerical assessment of these approximations.
from a statistical analysis of 300 000 fake objects, \cite{Delorme.these}
\footnote{http://tel.archives-ouvertes.fr/docs/00/35/10/10/PDF/these\_corrigee\_012009.pdf} derived the contamination percentage from a colour bin to another %(see Tab.~\ref{proba_conta}) 
and showed that the overall probability for any of the objects in the $1.3<i'-z'<1.7$ colour range to be falsely taken into account in our contamination calculation as a potential contaminant is $0.35\%$.

The number of contaminants in our 444 deg$^2$ sample is obtained in $z'$ magnitude bins of 0.1
mag and $i'-z'$ colour bins of 0.1 mag. The number of objects in these
small bins is usually smaller than 1. This number represents the
probability %$p_{ij}$ 
of contamination within one bin. %$(i,j)$. 
The total number of contaminants is 30, among which 23 have a resulting
colour $i'-z'<2.2$. 

\section{The brown dwarf space-density}
\label{lf}

The CFBDS patches contain a sufficient number of images and candidates
to provide significant statistics. Completeness and contamination are
thus handled separately in each patch before combining all data to derive the brown dwarf space-density.

In order to take into account the photometric errors, each object is spread over a colour range following the probability error distribution already used for contaminants (Fig.\ref{error}). In this process, we take into account that the true colour of the object is more likely to be bluer than redder, considering the density gradient as a function of $i'-z'$ colour: the probability shown in Fig.\ref{error} is weighted by the ratio of densities at the new colour and the observed colour, where densities as a function of colour are derived from our sample. Resulting objects with new $i'-z'$ colours and new $z'$ magnitudes beyond our selection limits are removed. 
Each object is thus split into several objects (about one hundred) whose weight is smaller than one, representing the actual likelihood that the object belongs to a given colour bin for a given observed colour.
%that may belong to different colour-magnitude bins $(i,j)$ than the initially observed colour.
The final weight $w$ is obtained by dividing by $c_{ij}$, the completeness in the  colour-magnitude bin $(i,j)$.
The contaminants are assigned a negative weight.

We determine the luminosity function of ultracool field dwarfs with the generalised form of the $V_{max}$ classical technique \citep{Schmidt.1968}, which allows to consider density gradients in the Galactic disc. 
We compute the maximum volume probed by our magnitude-limited survey at a given colour (that is at a given absolute magnitude). This geometric volume is corrected to take into account the decrease of the stellar density with increasing distance above the Galactic plane  \citep{Felten.1976,Stobie.1989,Tinney.1993}:
$$V_{Gen}=\Omega\frac{h^3}{\sin^3|b|}[2-(\xi^2+2\xi+2)\exp(-\xi)],$$
where $\Omega$ is the area of the survey, $b$ is the Galactic latitude of the field, $h$ is the thin disc scale height \citep[$h\sim250 pc$ from ][]{Robin.2003}, and $\xi=\frac{d\sin |b|}{h}$ with $d$ the maximum distance of detection.

For each object $d$ is estimated.
Thus the object is counted as the inverse of the maximum volume $V_{Gen}$ in which it is observed.
The luminosity function is the sum over all objects within an absolute magnitude bin: $$\Phi(M)=\sum\frac{w}{V_{Gen}},$$

where $w$ is the weight assigned to each object in this colour bin, taking into account the photometric errors and the completeness, as explained above.
 
The parallax is obtained from absolute magnitude - colour relations as described below.

\subsection{Absolute magnitude versus colours and spectral type relations}
\label{absmag}

\begin{figure*}[h]
\begin{minipage}{\textwidth}
\begin{center}
\includegraphics[scale=0.4,angle=270,clip=]{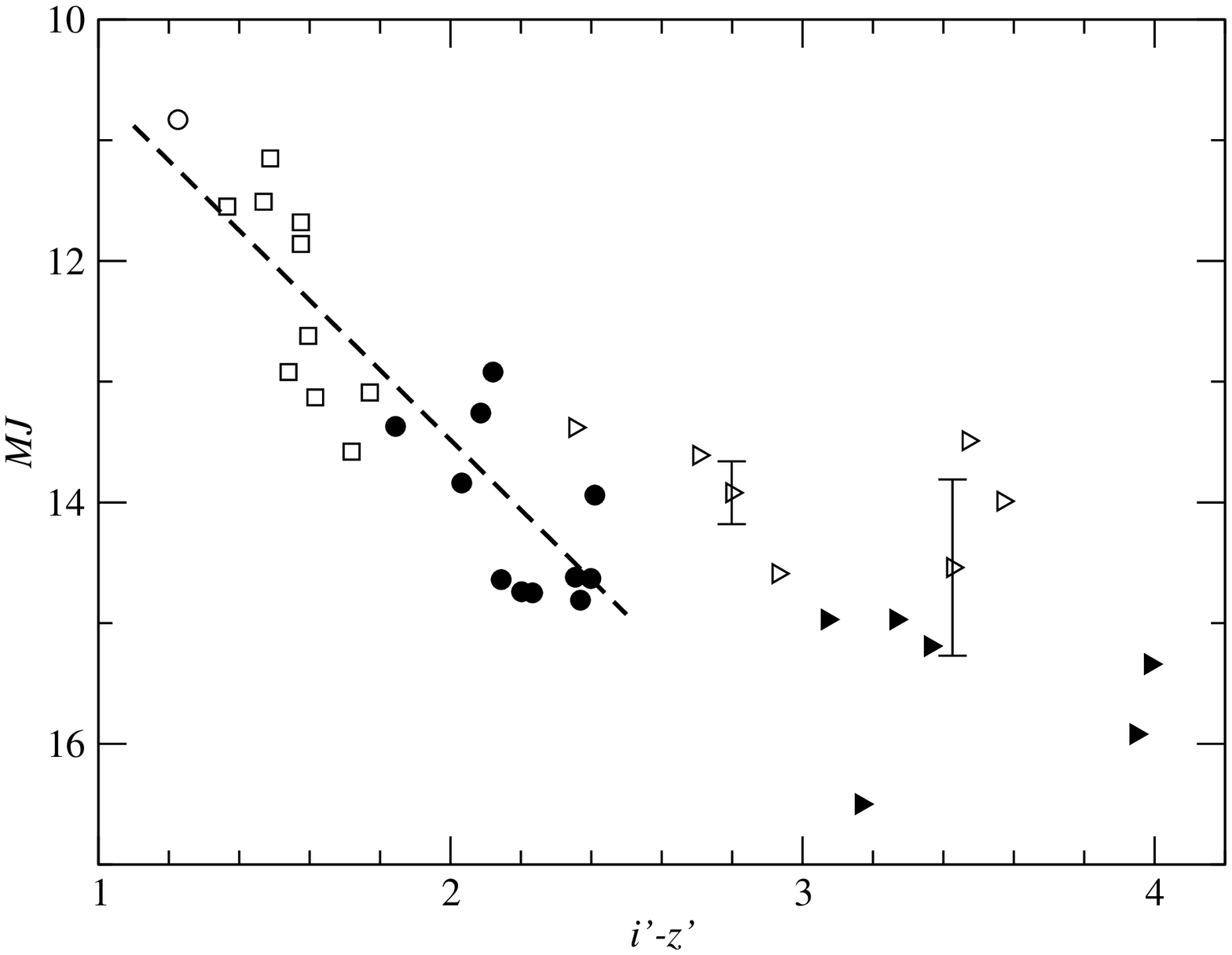}
\includegraphics[scale=0.4,angle=270,clip=]{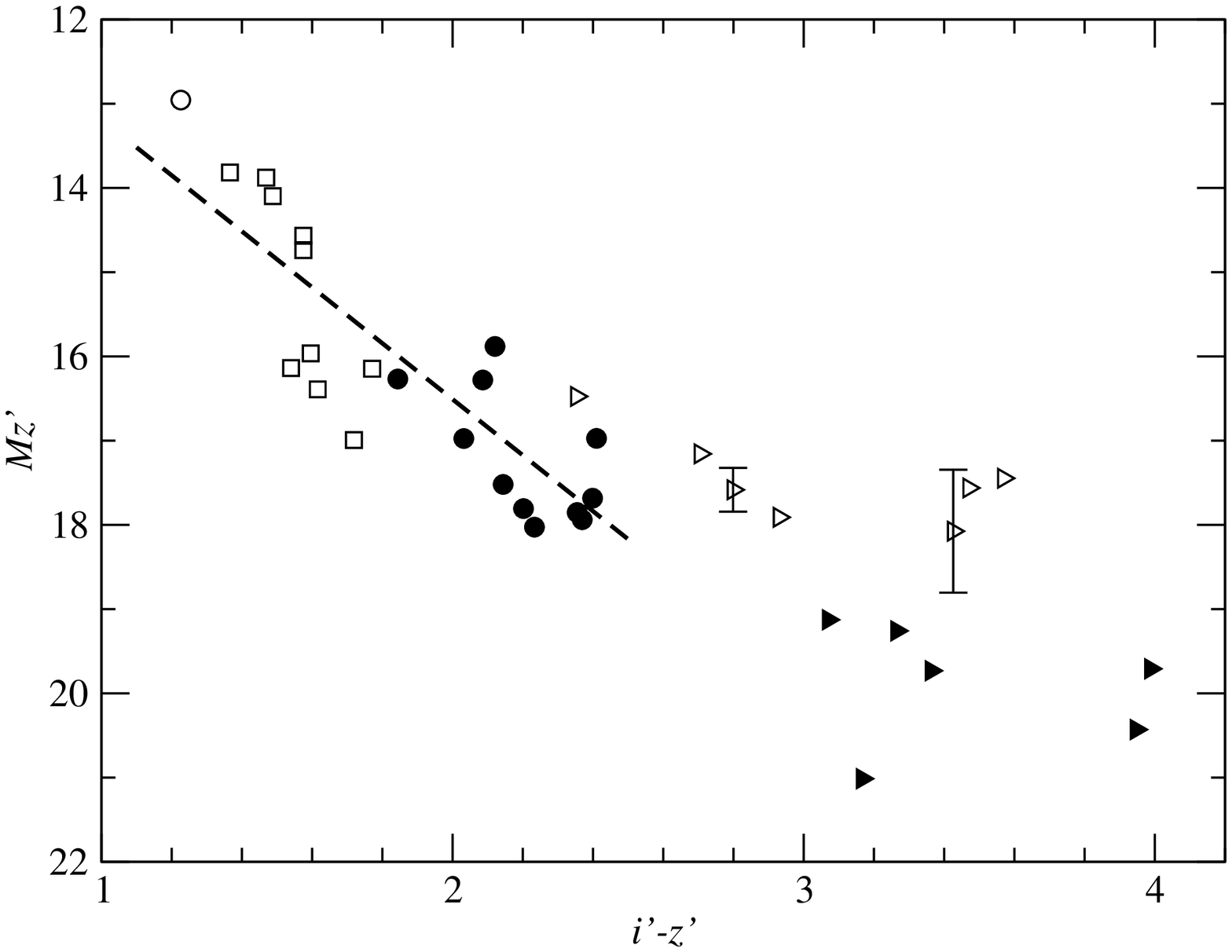}
\end{center}
\end{minipage}
\caption{
$M_J$ and $M_z'$ absolute magnitudes versus synthetic colour $i'-z'$ from available spectra of brown dwarfs with known parallax.  Open circles: dwarfs earlier than M8. Open squares: M8 to L4.5 dwarfs. Filled circles: L5 to L9.5 dwarfs. Open triangles: early T dwarfs. Filled triangles: late T dwarfs. The dotted line shows the derived absolute magnitude colour relation valid for late-M to L dwarfs. Error bars are indicated when larger than the symbol.
\label{magabscol}
}
\end{figure*}

\begin{figure*}[h]
\begin{minipage}{\textwidth}
\begin{center}
\includegraphics[scale=0.4,angle=270,clip=]{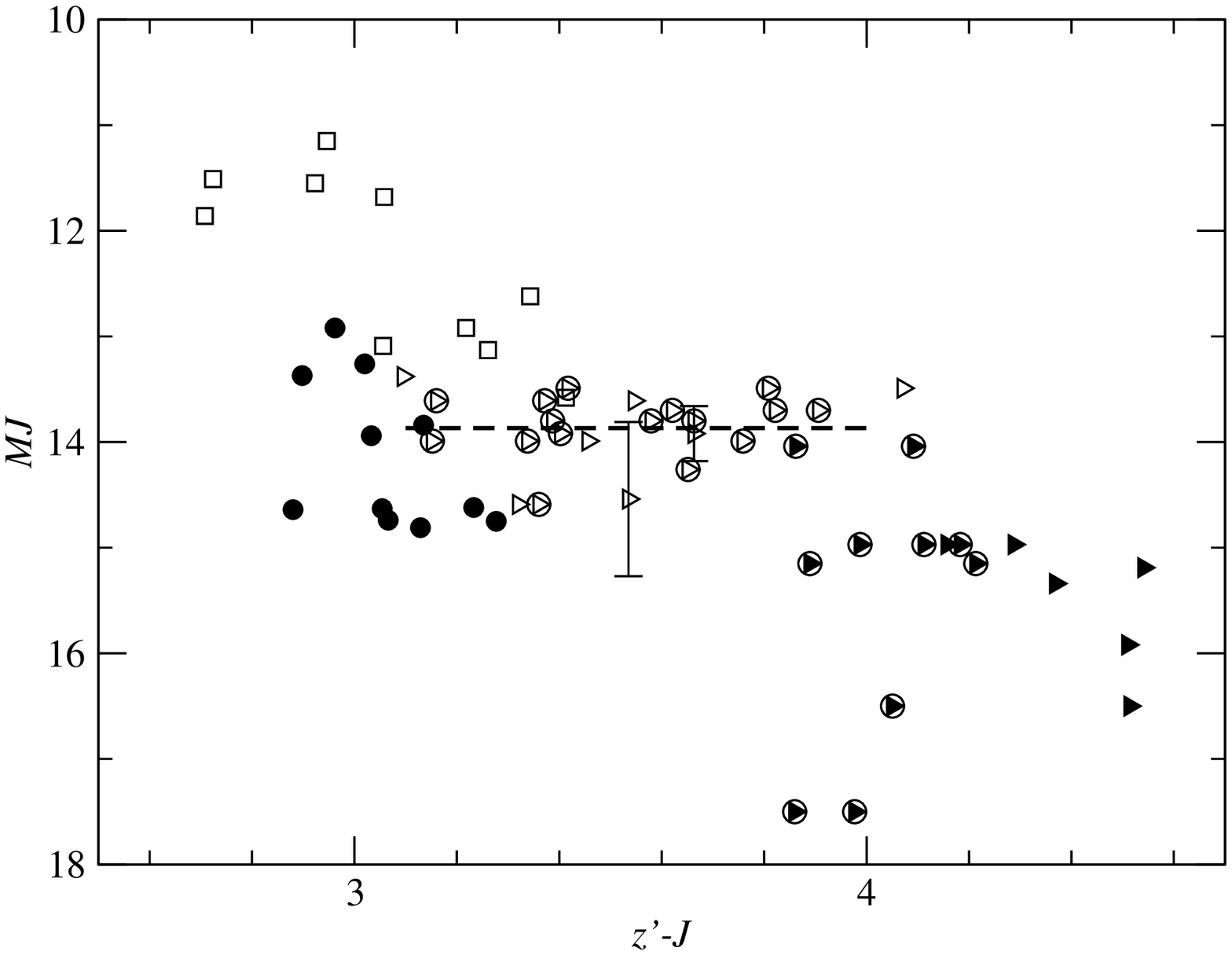}
\includegraphics[scale=0.4,angle=270,clip=]{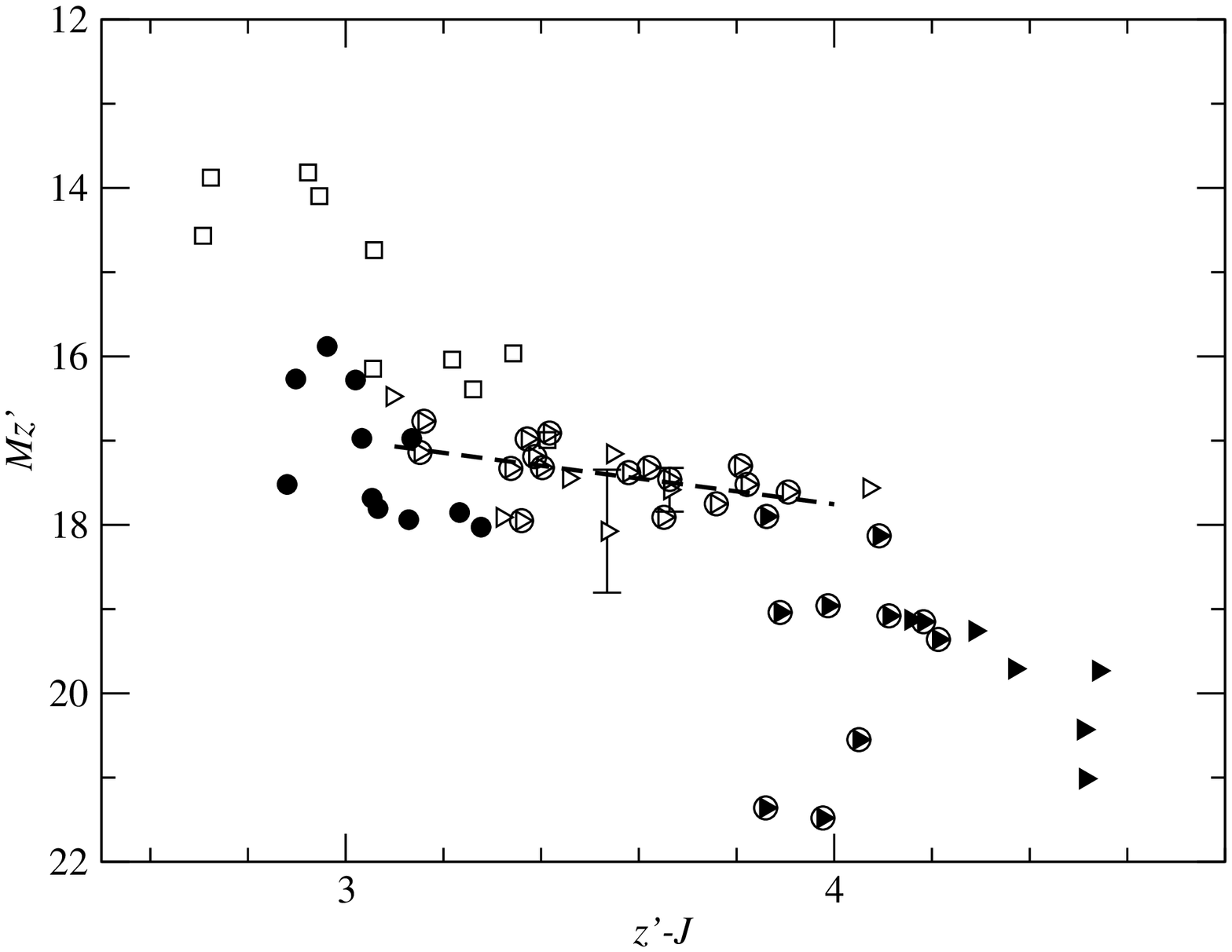}
\end{center}
\end{minipage}
\caption{
Same as Fig.~\ref{magabscol} for the $z'-J$ colour. Additional objects are shown (circled symbols). They are our candidates with spectroscopic follow-up. For these objects, $z'-J$ is the observed colour and the absolute magnitude is derived from their spectral type. The dotted line shows the derived absolute magnitude-colour relation valid for the early-T dwarfs. Error bars are indicated when larger than the symbol. Note that only objects with measured parallax -- non circled objects -- are used to derive the relation.
\label{magabscol2}
}
\end{figure*}

Among the publicly available spectra found in the L and T dwarf data
archive that are used to derive synthetic colours
(Sect.~\ref{classification}), 32 also have measured parallax. They allow
us to derive an absolute magnitude colour relation. The absolute
magnitude-colour diagrams are shown in 
Figs.~\ref{magabscol} and \ref{magabscol2}. The symbols indicate
the different spectral classes: open circle for a M7 dwarf, open
squares for M8 to L4.5 dwarfs,  filled circles for L5 to L9.5 dwarfs,
open triangles for T0 to T4.5 dwarfs and filled triangles for T5 and
later dwarfs. The dotted line shows the derived absolute magnitude - colour relation.
The scatter around the relations ranges from 0.2 to 0.8 mag with a mean value of 0.5 mag.

As one can see in the upper panels of Fig.~\ref{magabscol}, the
$i'-z'$ colour is a good luminosity estimator for late M and L
dwarfs, as shown before  in the SDSS photometric system \citep{West.2005}.
In the T dwarf domain, the $i'-z'$ colour is a poor
luminosity estimator. Moreover many T dwarfs are not detected in the
$i'$ band. Thus we use the $z'-J$ colour as a proxy. However the absolute $z'$ and $J$ band magnitudes of early T dwarfs are nearly constant. This translates into a flux increase in the $z'$ and $J$ bands from L8 to T4 spectral types \citep[known as $J$-band bump,][]{Dahn.2002,Tinney.2003,Vrba.2004}, contrarily to the classical decrease of luminosity with increasing spectral type. This behaviour at the L-T transition might be caused by the clearing of dust clouds in the atmosphere that reveals a deeper and higher temperature region of the atmosphere 
\citep{Burgasser.2002,Knapp.2004}. 

\begin{figure*}[h]
\begin{minipage}{\textwidth}
\begin{center}
\includegraphics[scale=0.4,angle=270,clip=]{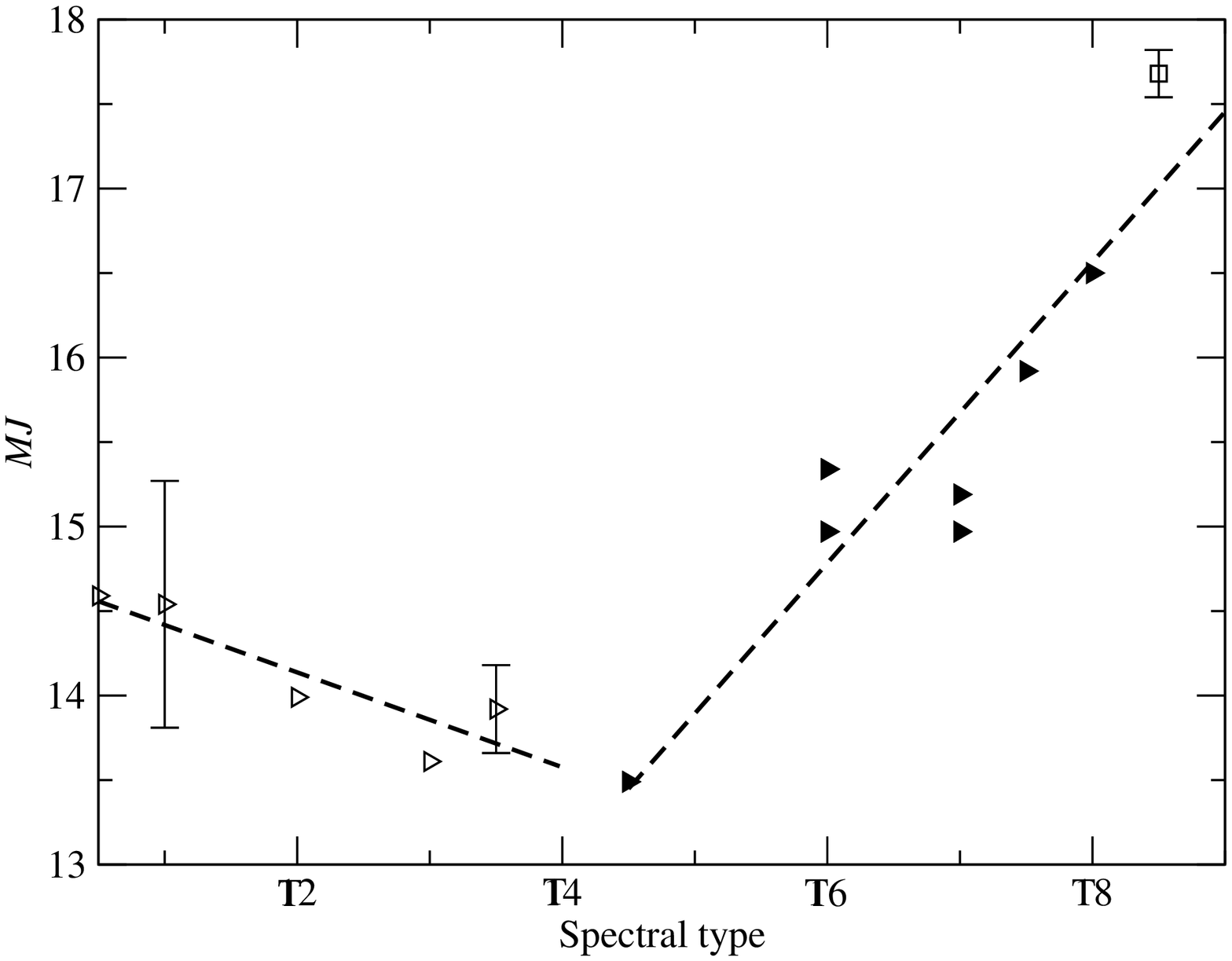}
\includegraphics[scale=0.4,angle=270,clip=]{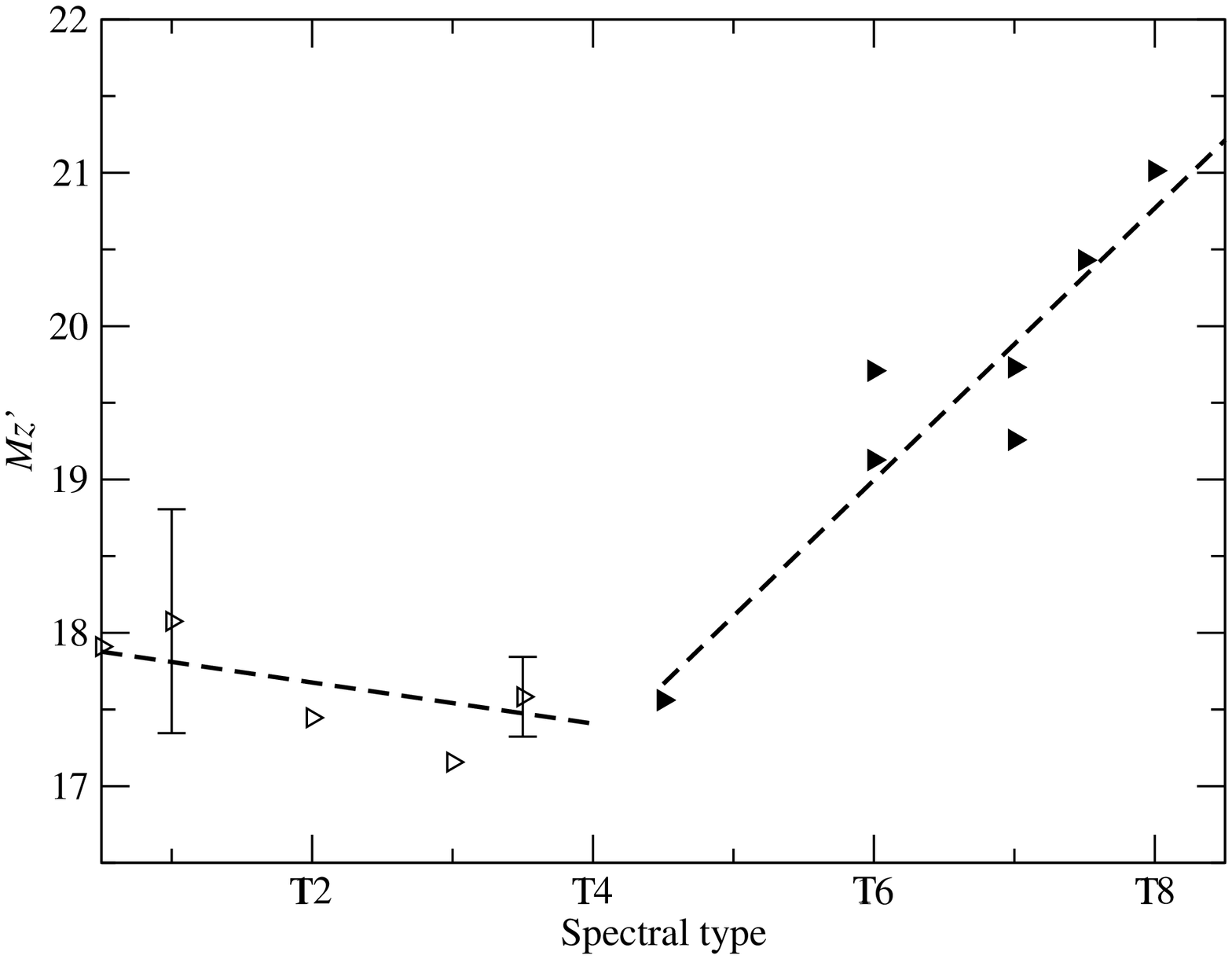}
\end{center}
\end{minipage}
\caption{
$M_J$ and $M_z'$ absolute magnitudes versus spectral type from available spectra of brown dwarfs with known parallax. The dotted line shows the derived absolute magnitude spectral type relation for the early-T and late-T dwarfs. Error bars are indicated when larger than the symbol. The open square with error bars shows the T8.5 companion to Wolf 940A, a M4 dwarf \citep{Burningham.2009}.
\label{magabsSpT}
}
\end{figure*}

Of our candidates, 32 were found to be T dwarfs from their spectroscopic
follow-up. They are plotted in Fig.~\ref{magabscol2} (circled
symbol) with their observed $z'-J$
colour and with $M_J$ and $M_z'$ determined from brown dwarfs with known
parallax that have the same spectral type. Whereas the agreement is good for early T dwarfs (open triangles),
the synthetic colours
of late T dwarfs may be biased and are redder than the observed colours of
T dwarfs with similar spectral type, as already discussed in Sect.~\ref{classification}, 
We performed spectroscopic
follow-up for all of our candidates with $z'-J > 4$. Thus we do not
use their $z'-J$ colour to derive their absolute magnitude but use their
spectral type instead.  Absolute magnitude versus spectral type relations are shown 
in Fig.~\ref{magabsSpT}. \\

To summarise, we use\\
-- $M_{z'}$ and $M_J$ vs $i'-z'$ relations for $i'-z'<2.4$ (L dwarfs):
$$M_{z'} = 3.32  (i-z') + 9.86; M_J = 2.90  (i-z') + 7.70$$
-- $M_{z'}$ and $M_J$ vs $z'-J$ relations for $i'-z'>2.4$ (early-T dwarfs):
$$M_{z'} = 0.74  (z'-J) + 14.86; M_J = 13.93$$
-- $M_{z'}$ and $M_J$ vs spectral type relations for dwarfs with spectroscopic follow-up (part of the early-T dwarfs and all late-T dwarfs).
$$\mbox{ for $\leq$T4: }M_J=-0.28*spT+17.49;  M_{z'}=M_{J}+(z'-J)$$
$$\mbox{ for $>$T4: }M_J=0.89*spT+0.55;  M_{z'}=M_{J}+(z'-J)$$

\subsection{Photometric distances}

\begin{figure}[h]
\begin{center}
\includegraphics[scale=0.4,angle=0,clip=]{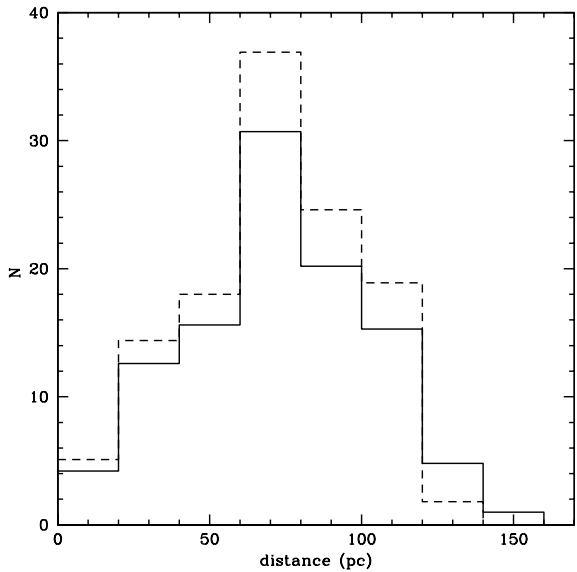}
\caption{
Photometric distance distribution computed in the $z'$ band of our sample,
without (solid line) and with (dotted line) contamination and completeness correction.
\label{dist}
}
\end{center}
\end{figure}

The previous relations are used to derive the photometric distances of the
brown dwarfs in our sample. The scatter in the absolute magnitude-colour relation translates to 
an uncertainty of 26\% on the distance estimate.
Figure~\ref{dist} shows the distribution of photometric distances
obtained from the $z'$ magnitudes 
without (solid line) and with (dotted line) contamination and
completeness correction of the sample. 
It mostly contains brown dwarfs within 20~pc and 120~pc, which most
probably belongs to the galactic disc, although it may contain close
subdwarfs that can be revealed by high kinematics or low-metallicity
features in the spectrum \citep{Burgasser.2002,Geballe.2002,Burgasser.2009,Burningham.2010b}. The distribution peaks between 60 and 100~pc,
which is about the maximal detection distance for late-L and mid-T
dwarfs. Most of the objects detected farther away than 120~pc are early T dwarfs brighter
than both late L and mid-T because of the $J$ and $z'$-bump.

\subsection{Malmquist bias}

The Malmquist bias \citep{Malmquist.1920} comes from the intrinsic dispersion in the absolute magnitude-colour relations 
and the limited-magnitude definition of the sample. A given colour (or spectral type) does not correspond to a unique luminosity, but rather a distribution due to intrinsic scatter in metallicity and age, and non detected binaries that appear brighter for their colour. This dispersion has two distinct effects:
\begin{itemize}
\item for a limited-magnitude sample, the mean absolute magnitude observed $\bar{M}$ is lower than the true mean magnitude $\bar{M_0}$ of objects with a given colour: among the most distant objects, the intrinsically most luminous ones are detected only (where the true intrinsic absolute magnitude $M_0 > M$),
\item the number of stars with given apparent magnitude and colour is larger, the brightest stars being observed at larger distances $d$ and the number of objects increasing as $d^3$.
\end{itemize}

\cite{Malmquist.1920} gave the first correction method for this bias, assuming a Gaussian distribution of the luminosity at a given colour, and later \cite{Stobie.1989} proposed the following analytic correction:
$$\frac{\Delta \Phi}{\Phi}\simeq(0.6ln10)^2\sigma^2-0.6ln10\sigma^2\frac{\Phi'}{\Phi},$$
where $\sigma$ is the luminosity dispersion, $\Phi$ the luminosity function, and $\Phi'=\frac{d\Phi}{dM}$. The first term is the volume element correction, the second term corrects for the magnitude translation from $M$ to ${M_0}$.

Because we already took into account the uncertainties in the colours due to photometric errors, $\sigma$ only depends on the scatter in the absolute magnitude-colour relation. This scatter ranges from 0.2 to 0.8 mag. We consider the mean value $\sigma=0.5$~mag.

\subsection{The luminosity function of systems}

Our magnitude-limited sample is also biased by binarity effect. An object in a non-resolved binary system appears brighter and enters our sample contrarily to the same isolated object. This also affects the colour of the objects. 
Direct imaging surveys \citep{Gizis.2003,Bouy.2003} that probe separations
of $>$2~AU show that $\sim10-15$\% of field brown dwarfs are binaries
with separations that peak between 2 and 4 AU. Lower separations probed by
spectroscopic surveys (see \cite{Joergens.2008} and references therein)
contribute a further $7_{-3}^{+5}$\% (separations $<0.3$~AU) or
$10_{-8}^{+18}$\% (separations $<3$~AU). The true BD binary fraction is
likely between 15 and 25\%, similar to estimates from \cite{Basri.2006}.

The purely observational data are not the luminosity function but a
stellar count as a function of colour and magnitude. To obtain the
luminosity function of the systems a first transformation is done 
with colour-magnitude or spectral type-magnitude relations. Bias could
be introduced during this first step, but the relations used (and their
limitations) are relatively well known. So the luminosity function of the
systems is a good proxy to compare stellar counts of different surveys
(with different sets of filters and magnitude limits). Moreover, it is
the luminosity function of systems which is used for estimating the
number of sources in any survey (with different seeing and spatial resolution).

As explained above, the correction of this function in a true
luminosity function (density of stars by magnitude bin) requires the
knowledge of the statistical multiplicity (fraction of binaries,
triple and higher order systems unresolved and distribution of mass
ratios between the components). To date such statistics for brown
dwarfs have still a significant margin of error.
Hence in this paper we chose to present the luminosity function of systems 
defined as the luminosity function uncorrected from the bias due to multiple stars.

\begin{figure*}[h]
\begin{center}
\includegraphics[scale=0.4,angle=0,clip=]{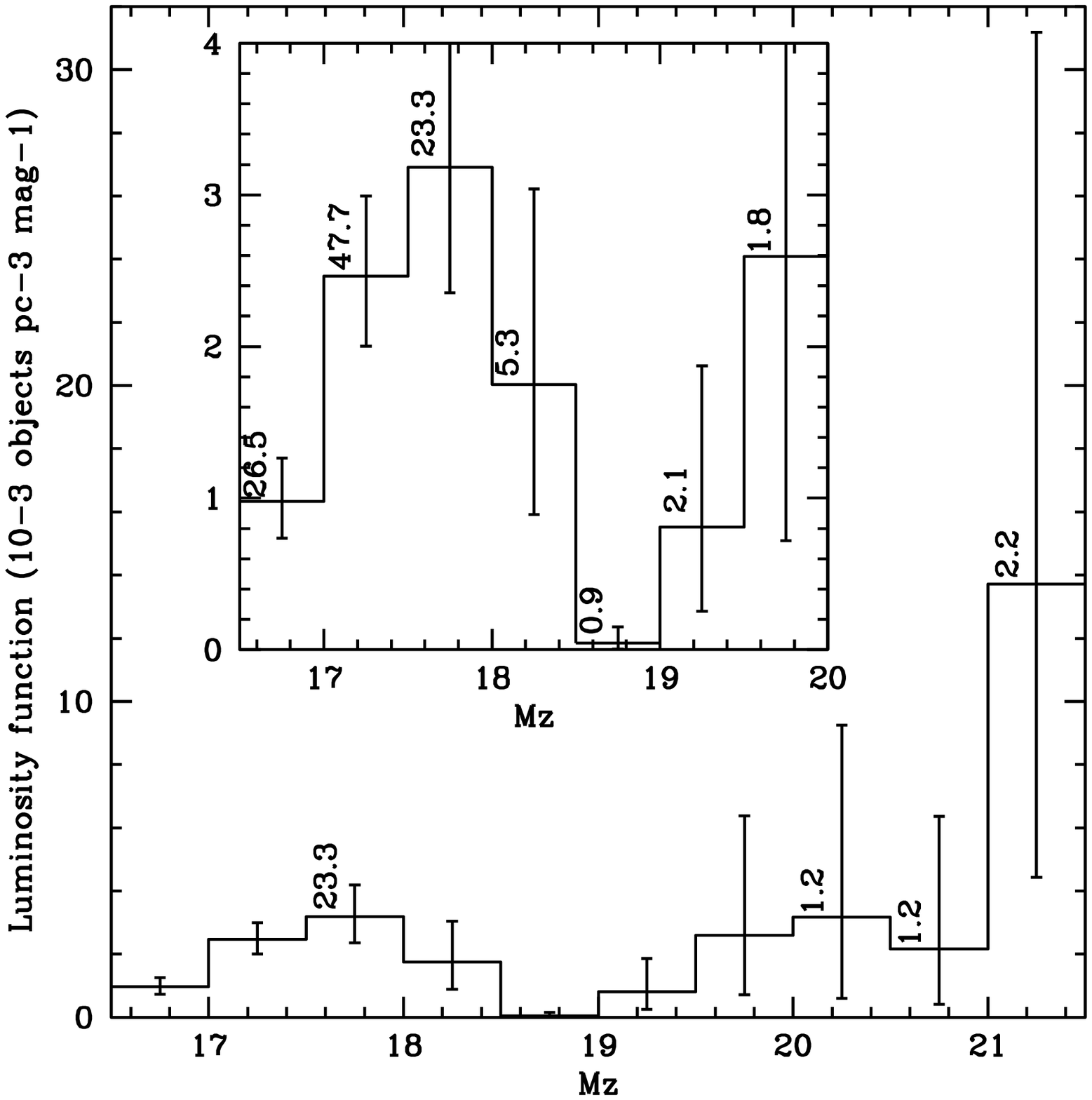}
\includegraphics[scale=0.4,angle=0,clip=]{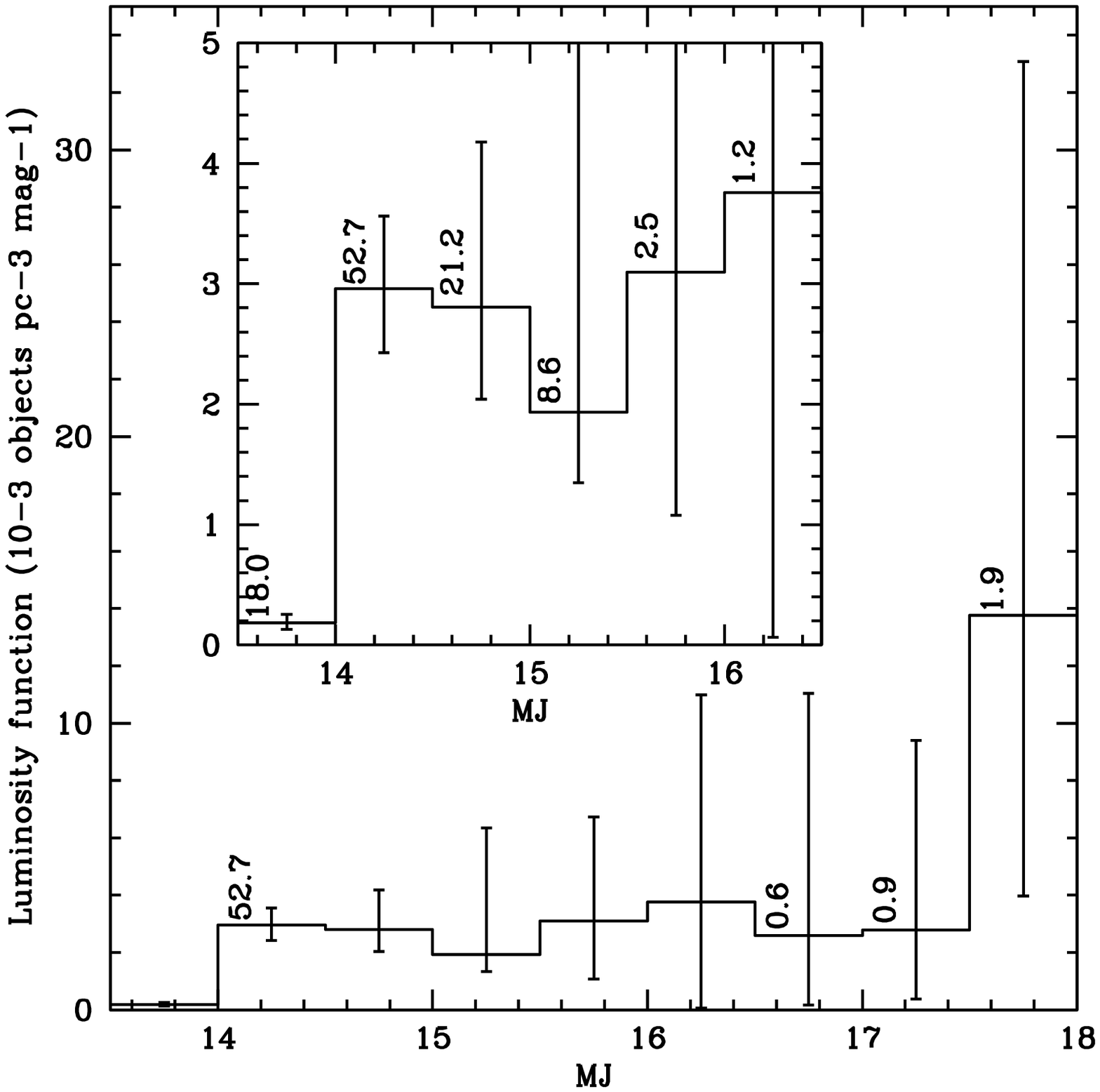}
\caption{
Luminosity function of systems computed in the $z'$ band (left) and the $J$ band (right) of our sample. \label{figfl}
}
\end{center}
\end{figure*}

Figure~\ref{figfl} shows the luminosity function of systems in the $z'$ and $J$ bands obtained from the brown dwarf sample corrected from contamination and completeness. The number of objects in magnitude bins are also indicated. The values are listed in Table~\ref{tablf}. Error bars represent the 95\% confidence level interval, derived from Bayesian statistics assuming a Poissonian distribution and a conjugate prior following the method described by \cite{Metchev.2008}.

\begin{table}[h]
\caption{Brown dwarfs $J-$band and $z'$-band luminosity function $\Phi$ ($10^{-3}$ objects pc$^{-3}$ mag$^{-1}$). Error bars give the 95\% confidence level interval, derived from Bayesian statistics. 
}             
\label{tablf}      
\centering                          
\begin{tabular}{c c c }        
\hline\hline                 
$M_J$ & $N_{tot}$ & $\Phi$ \\    
\hline                        
%13.25   &3.2                          & 0.17$^{+0.17}_{-0.10}$   \smallskip \\     
13.75   &18.0                        &$>$0.78$^{+0.30}_{-0.24}$     \smallskip \\            
14.25   &52.7                        &2.95$^{+0.60}_{-0.53}$     \smallskip   \\
14.75   &21.2                        &2.80$^{+0.93}_{-0.76}$     \smallskip  \\    
15.25   &8.6                          &1.93$^{+1.06}_{-0.78}$   \smallskip  \\
15.75   &2.5                          &3.09$^{+3.63}_{-2.02}$     \smallskip     \\ 
16.25   &1.2                          &3.75$^{+7.23}_{-3.04}$	      \smallskip  \\  
16.75   &0.6                          &2.59$^{+8.44}_{-2.42}$      \smallskip  \\
17.25   &0.9                          &2.78$^{+6.62}_{-2.41}$       \smallskip     \\
17.75   &1.9                          &13.75$^{+19.33}_{-9.77}$       \smallskip     \\
\hline                                  
\hline                                  
&&   \smallskip  \\
\end{tabular}
\hspace{0.5cm}\begin{tabular}{c c c }        
\hline\hline                 
$M_{z'}$ & $N_{tot}$ & $\Phi$ \\    
\hline                        
16.75   &26.5                        &0.98$^{+0.29}_{-0.24}$     \smallskip \\            
17.25   &47.7                        &2.46$^{+0.52}_{-0.46}$     \smallskip   \\
17.75   &23.3                        &3.18$^{+1.01}_{-0.83}$     \smallskip  \\    
18.25   &5.3                          &1.75$^{+1.29}_{-0.86}$   \smallskip  \\
18.75   &0.9                          &0.04$^{+0.11}_{-0.04}$     \smallskip     \\ 
19.25   &2.1                          &0.81$^{+1.06}_{-0.55}$	      \smallskip  \\  
19.75   &1.8                          &2.59$^{+3.78}_{-1.87}$      \smallskip  \\
20.25   &1.2                          &3.16$^{+6.08}_{-2.55}$       \smallskip     \\
20.75   &1.2                          &2.17$^{+4.18}_{-1.76}$       \smallskip     \\
21.25   &2.2                          &13.72$^{+17.46}_{-9.29}$   \smallskip \\     
\hline                                  
\hline                                  
\end{tabular}
\end{table}

One notices an increase at the faintest absolute magnitudes, due to T8 and later dwarfs. However, the uncertainties are very large, based on three objects only and the luminosity function remains consistently flat. Furthermore, the luminosity-spectral type relation used for distance estimate is rather rough (see Fig.~\ref{magabsSpT}). Assuming a luminosity 0.5 mag brighter leads to a luminosity function twice smaller in these bins. However, the T8.5 dwarf found around a M4 dwarf with measured parallax \citep[][ open square in Fig.~\ref{magabsSpT}]{Burningham.2009} tends to show that the late-T dwarfs luminosity is not underestimated, on the contrary. This uptick in number count is expected from luminosity function simulations \citep[see][and the pile-up of objects with $T_{eff}<500$ K in their Fig. 2]{Allen.2005}.

Finding half of the known ultracool brown dwarfs ($>$T8) could have been a statistical fluke. Future ultracool brown dwarf discoveries will build up this statistic and check whether this significant increase at the lower end of the  luminosity function that we find here is real or not. If true, this would mean either that the lowest mass brown dwarfs are more numerous than expected -- which would be different to what is observed in young cluster \citep[see e.g.][]{Moraux.2007} -- or that the number of old brown dwarfs of any mass is high. Indeed, whatever their mass, brown dwarfs cool 
down to the latest and coldest end of the spectral range after a few Gyr and should 
accumulate there because the cooling rate significantly slows down at these low 
temperatures. This speculation would agree interestingly with the hints from \cite{Gould.2009} that old-population brown dwarfs could be much more numerous than young ones, based on their a priori low probability detection of a thick-disk brown dwarf in a microlensing event. 

\section{Discussion}
\label{disc}

Table~\ref{lfcomp} summarises the comparison with other brown dwarf space densities in $M_J$ bins: the space density from \cite{Cruz.2007} computed from a 20 pc volume-limited sample of M7 to L8 dwarfs and the one obtained \cite{Allen.2005} from a sample of 14 T-dwarfs and a volume-limited sample of late M and L dwarfs.
As seen in Fig.~\ref{iz-zJ_synth}, the T dwarfs can be selected from their $i'-z'$ colours larger than $2.6$. For objects with no $i'$ detection, the $z'-J>3.5$ limit is also a good selection criterion for T dwarfs. Thus it is possible to compute the space density for T-dwarfs in different spectral type ranges and to compare with \cite{Metchev.2008} result . 
Our results agree with others. For early-T-dwarfs, they agree with the lower value found by \cite{Metchev.2008}.

%\begin{table}[h]
%\caption{Comparison of the brown dwarfs space densities $\rho$ (10$^{-3}$ objects pc$^{-3}$) obtained from CFBDS, and \cite{Cruz.2007,Metchev.2008}.}             
%\label{lfcomp}      
%\centering                          
%\begin{tabular}{c c c c c }        
%\hline\hline                 
%Spectral type &L5-T0 &T0.5-T5.5 &T6-T8 &T8.5-T/Y\\    
%\hline                        
%$\rho$(Metchev \& Cruz) &$> 1.5^{+}_{}0.2$ &$2.3^{+}_{}0.9$ &$4.7^{+}_{}3.0$ &---\\
%$\rho$(CFBDS) &$1.3^{+}_{}0.2$ &$1.6^{+}_{}0.2$ &$2.1^{+}_{}1.2$ &$12.2^{+}_{}7.3$\\
%\hline                                  
%\hline                                  
%\end{tabular}
%\end{table}

\begin{table*}[h]
\caption{Comparison of the brown dwarf space densities $\rho$ (10$^{-3}$ objects pc$^{-3}$) obtained from CFBDS given separately for L and T-dwarfs, and \cite{Allen.2005,Cruz.2007,Metchev.2008}.}             
\label{lfcomp}      
\centering                          
\begin{tabular}{c c c c c c c}        
\hline\hline                 
$M_J$ & $\rho$(Cruz) & $\rho$(Allen) & $\rho$(Metchev) &$\rho_L$(CFBDS) &$\rho_T$(CFBDS) &$\rho_{tot}$(CFBDS)\\    
%Spectral type & $\rho$(Cruz\&Metchev) &$\rho$(CFBDS)\\    
\hline                        
13.75 &$0.5\pm0.2$             &$1.0\pm0.3$             &--     	&-- 					   &$0.78^{+0.30}_{-0.24}$ \smallskip \\            
14.25 &$>0.7\pm0.2^a$      &$>1.7\pm1.4^a$         &--      &1.09$^{+0.30}_{-0.26}$     &1.86$^{+0.60}_{-0.49}$  \smallskip   \\
14.75 &$>0.3\pm0.2^a$      &$>2.3\pm1.4^a$          &--     &2.18$^{+0.78}_{-0.63}$      &0.62$^{+0.76}_{-0.41}$ \smallskip  \\    
15.25   &--                          &$>1.9\pm0.9^a$	&--    		&0.79$^{+0.49}_{-0.34}$     &1.14$^{+1.12}_{-0.67}$\smallskip  \\
15.75   &--                          &$2.0\pm1.5$	     &--       		&0.57$^{+1.63}_{-0.52}$      &2.52$^{+3.67}_{-1.82}$\smallskip     \\ 
16.25   &--                          &$4.7\pm0.3$      &--    		&--	     			             &3.75$^{+7.23}_{-3.04}$\smallskip  \\  
T0-T5.5 &--&--&$2.3^{+0.9}_{-0.9}$ &-- &$1.4^{+0.3}_{-0.2}$ \smallskip\\
T6-T8  &--&--&$4.7^{+3.1}_{-2.8}$  &-- &$5.3^{+3.1}_{-2.2}$ \smallskip \\
T8.5-T/Y &--&--&--     &-- &$8.3^{+9.0}_{-5.1}$ \smallskip\\
%16.75   &--                          & &2.59$^{+8.44}_{-2.42}$      \smallskip  \\
%17.25   &--                          & &2.78$^{+6.62}_{-2.41}$       \smallskip     \\
%17.75   &--                          & &13.75$^{+19.33}_{-9.77}$       \smallskip     \\
\hline                
\hline                
\end{tabular}

$^a$ a lower limit is given when the authors noted the incompleteness of their sample due to colour selection biases.

\end{table*}

The most important aim in deriving the luminosity function
for ultracool dwarfs is setting constraints on the mass function.
The computation of mass function from the luminosity function is quite straightforward for stars whose luminosity follows a simple function of their mass. The case is more complex for brown dwarfs that undergo continuous cooling and gravitational contraction during their life. The decrease of luminosity with time provokes a degeneracy: a low-mass brown dwarf can be as bright as a higher mass brown dwarf if younger. The only way to determine age and mass is to compare the spectra with theoretical atmosphere models, which still give quite uncertain results, because the physics of these cool atmospheres is very complex and not yet totally understood.
The non-relation between mass and luminosity makes difficult the determination of the mass function. A detailed determination of the substellar mass function is beyond the scop of this paper.

However, first estimates of the substellar mass function have been obtained in the L-dwarf domain: \cite{Reid.1999} with $1<\alpha<2$, \cite{Burgasser.2002PhD} $0.5<\alpha<1$, \cite{Allen.2005} $\alpha<0.3\pm0.6$, where $\alpha$ is the slope of the power-law mass function $\Psi(M ) = dN/dM \propto M^{-\alpha}$. These results are consistent with the substellar mass function in young open clusters showing $\alpha \simeq 0.6$ \citep[e.g.][]{Caballero.2007,Moraux.2007,Luhman.2007a}. Recently, \cite{Metchev.2008} derived a T dwarf space density that was mostly consistent with $\alpha = 0$ and the comparison of the observed number of T4-T8.5 dwarfs by \cite{Pinfield.2008} favoured $\alpha<0$. The analysis of 47 T-dwarfs found in the Large Area Survey (LAS) of UKIDSS also suggests that the substellar mass function is declining at lower masses \citep{Burningham.2010}.

\cite{Burgasser.2007b} performed simulations assuming different mass functions. The comparison between our luminosity function and these simulations is shown in Table~\ref{mfcomp}. It suggests that the best agreement is obtained with a flat mass function ($\alpha=0$) or even a decreasing mass function. Note that our values disagree for early T-dwarfs. However, the physics at the L/T transition is difficult to model, not only due to the cooling of the atmosphere but also to the clearing of the atmosphere that needs to model properly the hydrodynamics and the clouds formation \citep{Freytag.2010}, making the timescales uncertain. If confirmed, this low value would back \cite{Burgasser.2007} suggestion that the L/T transition occurs rapidly.

\cite{Pinfield.2008} suggested that a single power-law exponent is not optimal when describing the substellar mass function in the field, or that the different measured $\alpha$ values result from different models used to convert between mass and magnitude. Further studies based on more extended samples, such as the one we will be able to define at the completion of the CFBDS survey, or from the ongoing Large Area Survey performed by UKIRT, are needed to make reliable investigations on the mass function of field brown dwarfs.

\begin{table}[h]
\caption{Comparison of the brown dwarf space densities $\rho$ (10$^{-3}$ objects pc$^{-3}$) obtained by \cite{Burgasser.2007b} from simulations assuming different mass functions: $\Psi(M ) = dN/dM \propto M^{-\alpha}$.}             
\label{mfcomp}      
\centering                          
\begin{tabular}{c c c c c}        
\hline                
\hline                
Spectral type &$\rho$($\alpha=0$) &$\rho$($\alpha=0.5$)&$\rho$($\alpha=1$) &$\rho$(CFBDS)\\    
%Spectral type & $\rho$(Cruz\&Metchev) &$\rho$(CFBDS)\\    
\hline                        
L5-L9.5 &2.3 &2.9 &4.9 &$2.0^{+0.8}_{-0.7}$ \smallskip \\
T0-T5.5 &3.3 &4.4& 6.1 &$1.4^{+0.3}_{-0.2}$\smallskip\\
T6-T8  &6.4 &9.5 &14.4 &$5.3^{+3.1}_{-2.2}$\smallskip \\
\hline                                  
\hline                                  
\end{tabular}
\end{table}

\section{Conclusions}
\label{ccl}

At mid-course of the CFBDS, we define a uniform sample to investigate the field brown dwarf luminosity-function. We obtained spectroscopic follow-up for most of T-dwarf candidates. Because it is not realistic to obtain spectra for all L-dwarf candidates, we used the noise properties of the images to compute the contamination and completeness with good enough accuracy to properly characterise the sample. The sample covers a large spectral type range, including the L-T transition. It contains 56 L-dwarfs cooler than L5 and 41 dwarfs along the whole T-dwarfs sequence. We have measured the density of late L5 to T0 dwarfs to be $2.0^{+0.8}_{-0.7} \times 10^{-3}$ objects pc$^{-3}$, the density of T0.5 to T5.5 dwarfs to be $1.4^{+0.3}_{-0.2} \times 10^{-3}$ objects pc$^{-3}$, and the density of T6 to T8 dwarfs to be $5.3^{+3.1}_{-2.2} \times 10^{-3}$ objects pc$^{-3}$. Three latest dwarfs at the boundary between T and Y dwarfs give the high density $8.3^{+9.0}_{-5.1} \times 10^{-3}$ objects pc$^{-3}$, although the uncertainties are very large and the luminosity function is still consistent with being flat. 
Since then, at least three ultracool brown dwarfs ($>$T8) have also been discovered in UKIDSS \citep{Burningham.2008,Burningham.2009} that currently probes a slightly higher but comparable volume for late-T dwarfs. Even if these combined statistics still deal with a low number of objects, this could indicate that the number of ultracool brown dwarfs found by the CFBDS is not entirely due to a fluke of statistics and that many of these objects remain to be discovered in the solar neighbourhood. This can be expected as, whatever their mass, brown dwarfs cool down and finally accumulate in this spectral type range.

%Since then, another T9 dwarf, one T8.5 dwarf, and one T8.5 dwarf companion to a M4 dwarf have been found in UKIDSS \citep{Burningham.2008,Burningham.2009}.} UKIDSS is a survey which currently probes a slightly higher but comparable volume for late T dwarfs. Even if these combined statistics still deal with a low number of objects, this could indicate that the strickingly high density of ultracool brown dwarfs found by the CFBDS is not entirely due to a fluke of statistics.

\begin{acknowledgements}
Financial support from the "Programme National de Physique
Stellaire'' (PNPS) of CNRS/INSU, France, is gratefully acknowledged. We thank the queue observers at CFHT who obtained data for this study. We also thanks the NTT and NOT support astronomers for their help during the observations which led to these results. This research has made use of Aladin, operated at CDS, Strasbourg.
\end{acknowledgements}

\bibliographystyle{aa}
\bibliography{bib}

\end{document}